\documentclass[11pt]{article}
\usepackage{axodraw}
\usepackage{epsfig}
\usepackage{amsfonts}
\usepackage{amsmath}
\usepackage{bbm}
\input pix.sty

\renewcommand{\eq}{eq.~}
\renewcommand{\eqs}{eqs.~}
\renewcommand{\se}{sec.~}
\renewcommand{\ses}{secs.~}
\renewcommand{\fig}{fig.~}
\renewcommand{\figs}{figs.~}

\newcommand{\tinymsbar}{{\overline{\mbox{\tiny\rm{MS}}}}}
\newcommand{\Lambdamsbar}{{\Lambda_\tinymsbar}}

\newcommand{\Nf}{N_{\rm f}}
\newcommand{\Nc}{N_{\rm c}}

\newcommand{\Tc}{T_{\rm c}}
\newcommand{\mE}{m_\rmii{E}}
\newcommand{\gE}{g_\rmii{E}}
\newcommand{\gammaE}{\gamma_\rmii{E}}

\newcommand{\rmO}{{\mathcal{O}}}
\newcommand{\bmu}{\bar\mu}
\newcommand{\CA}{\Nc}

\def\lsi{\raise0.3ex\hbox{$<$\kern-0.75em\raise-1.1ex\hbox{$\sim$}}}
\def\gsi{\raise0.3ex\hbox{$>$\kern-0.75em\raise-1.1ex\hbox{$\sim$}}}
\newcommand{\lsim}{\mathop{\lsi}}
\newcommand{\gsim}{\mathop{\gsi}}

\newcommand{\fe}{\rmi{f}}
\newcommand{\bo}{\rmi{b}}

\newcommand{\rmii}[1]{{\mbox{\tiny\rm{#1}}}}
\newcommand{\re}{\mathop{\mbox{Re}}}

\newcommand{\Tint}[1]{{\hbox{$\sum$}\!\!\!\!\!\!\!\int\,}_{\!\!\!\!\raise-0.9ex\hbox{$\scriptstyle{#1}$}}}
\newcommand{\Tinti}[1]{{{\Sigma}\!\!\!\!\raise0.3ex\hbox{$\int$}_\rmii{${#1}$}}}

\newcommand{\NN}{{\mathbb{N}}}

\newcommand{\unit}{{\mathbbm{1}}} 
\newcommand{\bi}{\begin{itemize}}
\newcommand{\ei}{\end{itemize}}

\newcommand{\hide}[1]{ }

\def\TAsc(#1,#2)(#3,#4,#5)%
{\SetWidth{2.0}\CArc(#1,#2)(#3,#4,#5)\SetWidth{1.0}}
\def\Lwidth{3}

\def\TAgl(#1,#2)(#3,#4,#5){\SetWidth{2.0}\PhotonArc(#1,#2)(#3,#4,#5){\Lwidth}%
{6.283 #3 mul 360 div #4 #5 sub #4 #5 sub mul sqrt mul Tdensity mul}%
\SetWidth{1.0}}
\def\TLgl(#1,#2)(#3,#4){\SetWidth{2.0}\Photon(#1,#2)(#3,#4){\Lwidth}
{#1 #3 sub #1 #3 sub mul #2 #4 sub #2 #4 sub mul add sqrt Tdensity mul}%
\SetWidth{1.0}}
\newcommand{\piC}[1]{\;\parbox[c]{40pt}{\begin{picture}(120,60)(0,-20)
\SetWidth{1.0}\SetScale{0.35} #1 \end{picture}}\; }
\renewcommand{\picc}[1]{\;\parbox[c]{60pt}{\begin{picture}(60,30)(0,0)
\SetWidth{1.0}\SetScale{1.3} #1 \end{picture}}\; }
\def\ConnectedA(#1,#2,#3){\piC{#1(60,-15)(75,34,146) #2(60,75)(75,214,326)%
 #3(60,60)(20,190,350)%
 \GBoxc(0,30)(10,10){1} \GBoxc(120,30)(10,10){1}%
  }}
\def\ConnectedB(#1,#2,#3){\piC{#1(60,-15)(75,34,146) #2(60,75)(75,214,326)%
 #3(60,60)(60,0)%
 \GBoxc(0,30)(10,10){1} \GBoxc(120,30)(10,10){1}%
  }}
\def\ConnectedC(#1,#2){\piC{#1(60,-15)(75,34,146) #2(60,75)(75,214,326)%
 \GBoxc(0,30)(10,10){1} \GBoxc(120,30)(10,10){1}%
  }}
\def\ConnectedD(#1,#2){\piC{#1(60,-15)(75,34,146) #2(60,75)(75,214,326)%
 \GBoxc(0,30)(10,10){1} \GBoxc(120,30)(10,10){1}%
 \SetWidth{2.0} 
 \Line(55,55)(65,65)%
 \Line(55,65)(65,55)
  }}
\def\Lwidth{1.3}

\newcommand{\picu}[1]{\;\parbox[c]{30pt}{\begin{picture}(30,60)(0,0)
\SetWidth{1.0}\SetScale{1.0} #1 \end{picture}}\; }
\def\PolyA{\picu{%
 \Lqu(10,0)(10,60)%
}}
\def\PolyB{\picu{%
 \Lqu(10,0)(10,60)%
 \Agl(10,30)(15,270,90)%
}}
\def\PolyC{\picu{%
 \Lqu(10,0)(10,60)%
 \Agl(10,15)(8,270,90)%
 \Agl(10,45)(8,270,90)%
}}
\def\PolyD{\picu{%
 \Lqu(10,0)(10,60)%
 \Agl(10,30)(8,270,90)%
 \Agl(10,30)(16,270,90)%
}}
\def\PolyE{\picu{%
 \Lqu(10,0)(10,60)%
 \Agl(10,35)(12,270,90)%
 \Agl(10,25)(12,-90,10)%
 \Agl(10,25)(12,40,90)%
}}
\def\PolyF{\picu{%
 \Lqu(10,0)(10,60)%
 \Agl(10,35)(15,270,90)%
 \Lgl(10,35)(25,35)%
}}
\def\PolyG{\picu{%
 \Lqu(10,0)(10,60)%
 \Agl(10,30)(15,270,90)%
 \GCirc(25,30){5}{0.5}
}}
\newcommand{\pich}[1]{\!\!\parbox[c]{60pt}{\begin{picture}(60,60)(-30,0)
\SetWidth{1.0}\SetScale{0.7} #1 \end{picture}}\!\! }
\def\CoulA{\pich{%
 \Laqu(-20,20)(-20,60)%
 \Line(-20,0)(-20,20)
 \Lqu(20,0)(20,40)%
 \Line(20,40)(20,60)%
 \Lgl(-20,30)(20,30)%
}}
\def\CoulB{\pich{%
 \Laqu(-20,0)(-20,40)%
 \Line(-20,40)(-20,60)
 \Lqu(20,0)(20,40)%
 \Line(20,40)(20,60)%
 \Lgl(-20,40)(20,40)%
 \Agl(20,20)(10,270,90)%
}}
\def\CoulC{\pich{%
 \Laqu(-20,0)(-20,40)%
 \Line(-20,40)(-20,60)
 \Lqu(20,0)(20,40)%
 \Line(20,40)(20,60)%
 \Lgl(-20,40)(20,40)%
 \Agl(20,40)(10,270,90)%
}}
\def\CoulD{\pich{%
 \Laqu(-20,0)(-20,60)%
 \Lqu(20,0)(20,60)%
 \Lgl(-20,20)(20,20)%
 \Lgl(-20,40)(20,40)%
}}
\def\CoulE{\pich{%
 \Laqu(-20,0)(-20,60)%
 \Lqu(20,0)(20,60)%
 \Lgl(-20,20)(20,40)%
 \Lgl(-20,40)(-5,32.5)%
 \Lgl(5,27.5)(20,20)%
}}
\def\CoulF{\pich{%
 \Laqu(-20,0)(-20,60)%
 \Lqu(20,0)(20,40)%
 \Line(20,40)(20,60)%
 \Lgl(-20,20)(0,30)%
 \Lgl(-20,40)(0,30)%
 \Lgl(0,30)(20,30)%
}}
\def\CoulG{\pich{%
 \Laqu(-20,20)(-20,60)%
 \Line(-20,0)(-20,20)
 \Lqu(20,0)(20,40)%
 \Line(20,40)(20,60)%
 \Lgl(-20,30)(20,30)%
 \GCirc(0,30){5}{0.5}
}}
\def\CyclA{\pich{%
 \Laqu(-20,20)(-20,60)%
 \Laqu(-20,60)(20,60)%
 \Line(-20,0)(-20,20)
 \Lqu(20,0)(20,40)%
 \Lqu(-20,0)(20,0)%
 \Line(20,40)(20,60)%
 \Lgl(-20,30)(20,30)%
 \Agl(0,60)(10,180,360)%
}}
\def\CyclB{\pich{%
 \Laqu(-20,20)(-20,60)%
 \Laqu(-20,60)(20,60)%
 \Line(-20,0)(-20,20)
 \Lqu(20,0)(20,40)%
 \Lqu(-20,0)(20,0)%
 \Line(20,40)(20,60)%
 \Lgl(-20,30)(20,30)%
 \Agl(0,0)(10,0,180)%
}}
\def\CyclC{\pich{%
 \Laqu(-20,20)(-20,60)%
 \Line(-20,0)(-20,20)
 \Laqu(-5,60)(20,60)%
 \Line(-20,60)(-5,60)
 \Lqu(20,0)(20,40)%
 \Line(20,40)(20,60)%
 \Lqu(-20,0)(5,0)%
 \Line(5,0)(20,0)%
 \Lgl(-20,30)(20,30)%
 \Lgl(0,0)(0,25)%
 \Lgl(0,35)(0,60)%
}}
\def\CyclD{\pich{%
 \Laqu(-20,20)(-20,60)%
 \Line(-20,0)(-20,20)
 \Laqu(-20,60)(20,60)%
 \Lqu(20,0)(20,40)%
 \Line(20,40)(20,60)%
 \Lqu(-20,0)(5,0)%
 \Line(5,0)(20,0)%
 \Lgl(-20,30)(20,30)%
 \Lgl(0,0)(0,30)%
}}
\def\CyclE{\pich{%
 \Laqu(-20,20)(-20,60)%
 \Line(-20,0)(-20,20)
 \Laqu(-5,60)(20,60)%
 \Line(-20,60)(-5,60)
 \Lqu(20,0)(20,40)%
 \Line(20,40)(20,60)%
 \Lqu(-20,0)(20,0)%
 \Lgl(-20,30)(20,30)%
 \Lgl(0,30)(0,60)%
}}

\makeatletter \@addtoreset{equation}{section} \makeatother
\renewcommand{\theequation}{\arabic{section}.\arabic{equation}}
\makeatletter
\renewcommand\section{\@startsection {section}{1}{\z@}%
                                   {-5.5ex \@plus -1ex \@minus -.2ex}
                                   {2.3ex \@plus.2ex}%
                                   {\normalfont\large\bfseries}}
\renewcommand\subsection{\@startsection{subsection}{2}{\z@}%
                                     {-3.25ex\@plus -1ex \@minus -.2ex}%
                                     {1.5ex \@plus .2ex}%
                                     {\normalfont\normalsize\bfseries}}
\renewcommand\thesection {\@arabic\c@section}
\renewcommand\thesubsection   {\thesection.\@arabic\c@subsection}
\renewcommand{\@seccntformat}[1]{%
\csname the#1\endcsname.\hspace{1.0em}}
\makeatother

\begin{document}

\flushbottom

\begin{titlepage}

\begin{flushright}
BI-TP 2009/20\\
\vspace*{1cm}
\end{flushright}
\begin{centering}
\vfill

{\Large{\bf
 Dimensionally regularized Polyakov loop correlators
 \\[2mm] 
 in hot QCD
}} 

\vspace{0.8cm}

Y.~Burnier$^{\rm a}$, 
M.~Laine$^{\rm a}$, 
M.~Veps\"al\"ainen$^{\rm b}$ 

\vspace{0.8cm}

$^\rmi{a}$%
{\em
Faculty of Physics, University of Bielefeld, 
D-33501 Bielefeld, Germany\\}

\vspace{0.3cm}

$^{\rm b}$%
{\em 
Department of Physics, 
P.O.Box 64, FI-00014 University of Helsinki, Finland\\}

\vspace*{0.8cm}

\mbox{\bf Abstract}
 
\end{centering}

\vspace*{0.3cm}
 
\noindent
A popular observable in finite-temperature lattice QCD is the so-called 
singlet quark--anti\-quark free energy, conventionally defined in Coulomb 
gauge. In an effort to interpret the existing numerical data on this 
observable, we compute it at order $\rmO(\alpha_s^2)$ in continuum, 
and analyze the result at various distance scales. At short 
distances ($r \ll 1/\pi T$) the behaviour matches that of the 
gauge-independent zero-temperature potential; on the other hand 
at large distances ($r \gg 1/\pi T$) the singlet free energy appears 
to have a gauge-fixing related power-law tail. At infinite
distance the result again becomes physical in the sense that 
it goes over to a gauge-independent disconnected
contribution, the square of the expectation value of the trace of 
the Polyakov loop; we recompute this quantity at $\rmO(\alpha_s^2)$, 
finding for pure SU($\Nc$) a different non-logarithmic term than 
in previous literature, and adding for full QCD the quark contribution.
We also discuss the value of the singlet free energy in a general 
covariant gauge, as well as the behaviour of the cyclic Wilson loop 
that is obtained if the singlet free energy is made gauge-independent
by inserting straight spacelike Wilson lines into the observable. 
Comparisons with lattice data are carried out where possible. 

\vfill

 
\vspace*{1cm}
  
\noindent
November 2009

\vfill

\end{titlepage}

%
\section{Introduction}

One of the classic probes for forming a thermalized partonic medium
in a heavy ion collision is the change that this should cause
in the properties of heavy quarkonium~\cite{ms}. 
It is traditional to address quarkonium through potential
models, but in the finite-temperature context this has been
hampered by the multitude of independent definitions of a potential 
that could in principle be introduced. A possible way to reduce  
the degeneracy is to use the weak-coupling
expansion as a test bench; at least within that approach, it should  
be possible to derive from QCD the potential that plays 
a physical role. Indeed, it has been discovered that the 
relevant potential is none of the plethora of conventional ones 
but a new object, which even has an imaginary part, representing 
a decoherence of the quark-antiquark state that is induced by the 
thermal medium~\cite{static}--\cite{nb3}. Subsequent to the perturbative 
introduction, it appears that it may be possible to promote
the corresponding definition to the non-perturbative level~\cite{akr}. 

In this paper, we do {\em not} discuss the question of  
a proper definition of a 
static potential at finite temperatures, but rather set 
a more modest goal. 
Extensive lattice measurements have already been carried out
on many potentials~(see, e.g., 
refs.~\cite{q_average}--\cite{whot} and references therein), and
first tests~\cite{akr} suggest that the real part of the proper 
potential may overlap with one of the existing objects,
namely the ``singlet'' quark-antiquark free energy. This concept is, however,
based on gauge fixing, more precisely on the Coulomb gauge
(cf.\ \eq\nr{Vsinglet} below). 
We are not aware of a satisfactory {\it a priori} theoretical justification 
for this choice; the original motivation appeared rather to be practical, 
in that the Coulomb gauge potential empirically  
reproduces the expected zero-temperature behaviour at short
distances (see, e.g., refs.~\cite{fz,whot}), 
and also displays manageable statistical fluctuations
as well as a good scaling with the volume 
and the lattice spacing~\cite{q_singlet}. 

Given these empirical observations, we feel that 
there may be room also for some theoretical work
of the Coulomb gauge potential. The purpose of this paper is 
straightforward: we compute the Coulomb gauge potential, 
and several variants thereof, at next-to-leading order 
in the weak-coupling expansion
(i.e.\ $\rmO(g^4)$, 
where $g^2 = 4\pi \alpha_s$ is the QCD coupling constant)
in dimensional regularization. 
The results turn out, indeed, 
to yield a qualitative surprise that could have a bearing 
on the interpretation of the Coulomb gauge lattice data.   

The paper is organized as follows. 
In \se\ref{se:pol} we compute the expectation value 
of a single Polyakov loop in dimensional regularization, 
and compare the result with literature. 
In \se\ref{se:Coulomb} we analyze the Coulomb gauge 
singlet quark--antiquark free energy 
at various distance scales, identifying  
both fortunate physical aspects as well as what looks like 
an unfortunate, if numerically small,  
gauge artifact in the result. 
In \se\ref{se:xi} we briefly discuss the 
corresponding object in a general covariant gauge, while 
\se\ref{se:cyclic} looks into the so-called
cyclic Wilson loop; this gauge-invariant completion 
removes the gauge artifact from the Coulomb gauge result 
but with the price of introducing new problems. 
Sec.~\ref{se:concl} summarizes our main findings.

%
\section{Polyakov loop expectation value}
\la{se:pol}

%
\subsection{Basic setup}

Employing conventions where the covariant derivative 
reads $D_\mu = \partial_\mu - i g_\rmii{B} A_\mu$, with
$g_\rmii{B}$ the bare gauge coupling and $A_\mu$ 
a traceless and hermitean gauge field, we define a ``Polyakov loop''
at spatial position $\vec{r}$ through
\be
 P_\vec{r} \equiv 
 \unit + ig_\rmii{B} \int_0^\beta\! {\rm d}\tau A_0(\tau,\vec{r}) 
 + (ig_\rmii{B})^2 \int_0^\beta \! {\rm d}\tau \, 
 \int_0^\tau \! {\rm d}\tau' \, 
 A_0(\tau,\vec{r}) A_0(\tau',\vec{r}) + \ldots 
 \;. \la{Pdef}
\ee
It transforms in a gauge transformation $U(\tau,\vec{r})$ as 
$P_\vec{r} \to U(\beta,\vec{r})\, P_\vec{r}\, U^{-1}(0,\vec{r})$. 
Likewise, for future reference, 
a straight spacelike Wilson line from origin to $\vec{r}$, 
at a fixed time coordinate $\tau$, is denoted by 
\be
 W_\tau = 
 \unit + ig_\rmii{B} \int_0^1 \! {\rm d}\lambda \, 
 \vec{r}\cdot \vec{A}(\tau,\lambda \vec{r})
 + (ig_\rmii{B})^2 
 \int_0^1 \! {\rm d}\lambda \, 
 \int_0^\lambda \! {\rm d}\lambda' \, 
 \vec{r}\cdot \vec{A}(\tau,\lambda \vec{r}) \,
 \vec{r}\cdot \vec{A}(\tau,\lambda' \vec{r})
 + \ldots
 \;, \la{Wdef}
\ee
and transforms as 
$W_\tau \to U(\tau,\vec{r})\, W_\tau \, U^{-1}(\tau,\vec{0})$. 
Note that because of the usual periodic boundary conditions
in the time direction, $W_\beta = W_0$.

In order to simplify the notation somewhat, we in general leave out 
the subscript from $g_\rmii{B}$ in the following formulae. It is to 
be understood that at the initial stages it is the bare 
coupling that is referred to, while at the end the 
bare coupling is re-expressed in terms of the renormalized
one, 
\be
 g_\rmii{B}^2 = g^2 + \frac{g^4 \mu^{-2\epsilon}}{(4\pi)^2} 
 \frac{2\Nf - 11\Nc}{3\epsilon} + \rmO(g^6)
 \;, \la{gBare}
\ee
where $g^2$ denotes the renormalized gauge coupling in the $\msbar$ scheme, 
with a scale parameter $\bmu^2 = 4 \pi e^{-\gammaE} \mu^2$.  The factor
$\mu^{-2\epsilon}$ is normally not displayed explicitly.

Within perturbation theory, assuming that the $Z(\Nc)$
center symmetry is broken in the trivial direction, either 
spontaneously or explicitly due to matter fields in the fundamental 
representation, we can then define the expectation value of the trace
of the Polyakov loop through~\cite{mls} 
\be
 \psi_\rmii{P} \equiv \frac{1}{\Nc} \langle \tr [ P_\vec{r} ]  \rangle
 \;, 
\ee 
where the brackets refer to averaging in a thermal 
ensemble at a temperature $T$. 
On the non-perturbative level, 
$\psi_\rmii{P}$ plays a role as 
the disconnected part of some 2-point function, e.g.\footnote{%
  The second equality 
  can be understood by writing $P_\vec{r} = 
  \fr{\unit}{\Nc} \tr[P_\vec{r}] + \hat P_\vec{r}$,  
  where $\hat P_\vec{r}$ is traceless and changes in gauge 
  transformations like an adjoint scalar; $\hat P_\vec{r}$
  should not have correlations at infinite distance, and 
  our results confirm this expectation at $\rmO(\alpha_s^2)$.   
  } 
\be
 |\psi_\rmii{P}|^2 
 = 
 \lim_{|\vec{r}|\to\infty}
 \frac{1}{\Nc^2}
 \langle \tr[P_\vec{r}] \, \tr[P^\dagger_\vec{0} ] \rangle  
 = 
 \lim_{|\vec{r}|\to\infty}
 \frac{1}{\Nc}
 \langle \tr[P_\vec{r} P^\dagger_\vec{0} ] \rangle_\rmi{Coulomb}
 \;.   \la{asympt}
\ee 

Any practical computation in finite-temperature QCD is hampered
by infrared divergences. There are two kinds: milder ones, associated
with colour-electric modes and the scale $gT$, which can be cured 
by a certain resummation of the perturbative series; and more serious 
ones, associated with colour-magnetic modes and the scale $g^2T/\pi$,
which can only be handled by a non-perturbative study of three-dimensional
pure Yang-Mills theory~\cite{linde}. For the observables that we are concerned
with, the first type plays a more prominent role;
it can be handled by writing the observable as 
\be
 \psi_\rmii{P} = \Bigl[ 
          (\psi_\rmii{P})_\rmii{QCD} - 
          (\psi_\rmii{P})_\rmii{EQCD} \Bigr]_\rmii{unresummed} + 
          \Bigl[ (\psi_\rmii{P})_\rmii{EQCD} \Bigr]_\rmii{resummed} 
 \;. \la{resum}
\ee 
The difference in the first square brackets is infrared finite (provided that
the correct low energy  effective  theory~\cite{dr}, called EQCD~\cite{bn}, 
is used), and can be computed in naive perturbation 
theory. The second term is infrared sensitive, 
but the resummations required for computing it can be implemented in 
the simple three-dimensional EQCD framework. 

%
\begin{figure}[t]
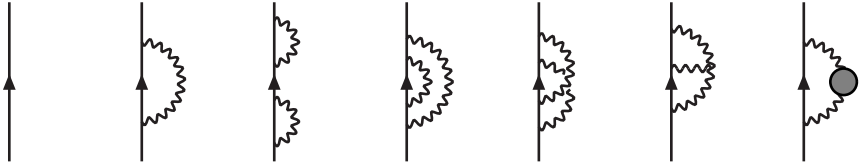


\begin{eqnarray*}
&& 
 \hspace*{-2cm}
 \PolyA \quad\; 
 \PolyB \quad\; 
 \PolyC \quad\; 
 \PolyD \quad\; 
 \PolyE \quad\; 
 \PolyF \quad\; 
 \PolyG \quad 
\end{eqnarray*}

\caption[a]{\small 
The graphs contributing to the expectation value of a single
Polyakov loop up to $\rmO(g^4)$, with the filled blob denoting 
the 1-loop gauge field self-energy.} 
\la{fig:Poly}
\end{figure}
%

The graphs contributing to the expectation 
value of a single Polyakov loop up to $\rmO(g^4)$
are shown in \fig\ref{fig:Poly}.
The computation of the graphs is in principle straightforward, however 
some care is needed when treating Matsubara zero modes. We separate their 
effects explicitly, writing
\ba
 \int_0^\tau \! {\rm d}\tau' \, 
 e^{i q_n \tau'}
 & = & 
  \delta_{q_n} \, \tau  
 + 
 (1-\delta_{q_n})
 \frac{1}{i q_n} \Bigl( e^{i q_n \tau} - 1 \Bigr)
 \;, \la{tri1} \\ 
 \int_0^\tau \! {\rm d}\tau' \, \tau' \, 
 e^{i q_n \tau'}
 & = & 
  \delta_{q_n} \, \frac{\tau^2}{2}  
 + 
 (1-\delta_{q_n})
 \biggl[ \frac{\tau}{i q_n} e^{i q_n \tau}
 + \frac{1}{q_n^2} \biggl( e^{i q_n \tau} - 1 \biggr)
 \biggr]
 \;, \la{tri2}
\ea
etc, where $q_n$ is a bosonic Matsubara mode, $q_n = 2 \pi T n$,  
and $\delta_{q_n} \equiv \delta_{n,0}$ is a Kronecker delta function. 
Adding up the graphs in any gauge where the gauge field propagator, 
$G_{\mu\nu}$, has the property $G_{0i}(0,\vec{k}) = 0$, 
the result (for the moment unresummed) 
can be written as\footnote{%
 Here and in the following we often use $\exp(x)$ as  
 a shorthand for $1 + x + x^2/2$; no proofs
 concerning the exponentiation of higher order corrections are provided. 
 }  
\ba
 \psi_\rmii{P}  \!\! & = & \!\! 
 \exp\biggl[ 
   - \fr12 g^2 C_F \beta \int_{\vec{k}} \! 
      G_{00}(0,\vec{k})
 \biggr] 
 + \fr12 g^2 C_F \beta \int_{\vec{k}} \! 
     G_{00}^2(0,\vec{k}) \, \Pi_{00}(0,\vec{k})
 \nn & & \; 
  - \fr12 g^4 C_F \CA  
   \int_{\vec{k},\vec{q}} \! 
    \biggl\{
     \frac{\beta^2 G_{00}(0,\vec{k}) G_{00}(0,\vec{q})}{24}
   \nn & & \; \hspace*{1.2cm}
     + \sum_{q_n'} 
     \Bigl[ - G_{00}(0,\vec{k}) + \fr12 G_{00}(q_n,\vec{k})
     \Bigr] \frac{G_{00}(q_n,\vec{q})}{q_n^2}
   \nn & & \; \hspace*{1.2cm}
    - 2  \sum_{q_n'} 
     G_{00}(0,\vec{k})
     \biggl[
        G_{00}(q_n,\vec{q}) \frac{k_i - q_i}{q_n} +  
        G_{0i}(q_n,\vec{q})
     \biggr] G_{0i}(q_n,\vec{q+k})
   \biggr\}
 \;, \la{psi_P_full}
\ea
where 
$C_F = (\Nc^2 - 1)/ 2 \Nc$;
$\beta \equiv 1/T$; 
$\Pi_{\mu\nu}$ is the gauge field self-energy;  
and $\sum_{q_n'} \equiv \sum_{q_n \neq 0}$.

Inserting the propagator of the general covariant gauge (\eq\nr{G_xi} below)
as well as the corresponding self-energy (\eq\nr{Pi_xi} below), it can be 
checked that the value of the expression in \eq\nr{psi_P_full} is 
gauge-parameter 
independent. The same result is also obtained in the Coulomb gauge, 
by making use of \eqs\nr{G_C}, \nr{Pi_C}. In explicit form, 
the gauge-independent part can be written as 
\ba
 \psi_\rmii{P} & = & 1 - \frac{g^2 C_F \beta}{2} \int_{\vec{k}} \frac{1}{k^2}
 + \fr12 \biggl( \frac{g^2 C_F \beta}{2} \int_{\vec{k}} \frac{1}{k^2} 
   \biggr)^2
 \nn & & \; 
 + \frac{g^4 C_F \Nf}{2} \sum_{ \{ q_n \} } \int_{\vec{k,q}}
 \biggl[ 
 -\frac{2}{k^4 Q^2} 
 + \frac{4 q_n^2}{k^4 Q^2(Q+K)^2}
 + \frac{1}{k^2 Q^2 (Q+K)^2}
 \biggr]
 \nn & & \; 
 - \frac{g^4 C_F \CA}{2} \sum_{ q_n' } \int_{\vec{k,q}}
 \biggl[ 
   \frac{2-D}{k^4 Q^2}
 + \frac{2 (D-2) q_n^2}{k^4 Q^2(Q+K)^2}
 + \frac{2 }{k^2 Q^2 (Q+K)^2}
 \nn & & \; \hspace*{3cm}
 + \frac{1}{2 q_n^2 Q^2 (q_n^2 + k^2)}
 - \frac{1}{k^2 q_n^2 Q^2}
 \biggr]
 \nn & & \; 
 - \frac{g^4 C_F \CA}{2} \int_{\vec{k,q}}
 \biggl[
   \frac{2-D}{k^4 q^2}
 + \frac{2 }{k^2 q^2 (q+k)^2}
 + \frac{\beta^2}{24 k^2 q^2}   
 \biggr]
 \;, \la{psi_P_expl}
\ea
where 
$k \equiv |\vec{k}|$, $q \equiv |\vec{q}|$,
$k+q \equiv |\vec{k+q}|$; 
$K\equiv (0,\vec{k})$, $Q \equiv (q_n,\vec{q})$; and 
we have separated the Matsubara zero mode contribution 
in the term of $\rmO(g^4)$.

%
\subsection{Soft-mode contribution}

It is immediately visible from \eq\nr{psi_P_expl} that
the $\vec{k}$-integral in the term of $\rmO(g^4)$
is infrared divergent, $\sim \int_{\vec{k}} 1/k^4$. The coefficient
of the divergence is, however, nothing but the Debye mass parameter: 
\ba
 \psi_\rmii{P} & = &  \ldots + \frac{g^4 C_F}{2} \int_{\vec{k}} \frac{1}{k^4}
 \biggl[ \int_{\vec{q}} \biggl( 2 \Nf \sum_{\{ q_n \} }
 -  \CA (D-2) \sum_{ q_n' } \biggr) 
 \biggl( - \frac{1}{Q^2} + \frac{2 q_n^2}{Q^4} \biggr)
 + \rmO(k^2) 
 \biggr]
 \nn & = &
 \ldots + \frac{g^2 C_F \beta}{2} \int_{\vec{k}} \frac{1}{k^4}
 \biggl[ \mE^2 
 + \rmO(k^2) 
 \biggr] 
 \;, 
\ea
where we made use of the sum-integrals in \eqs\nr{si1}, \nr{si2}, 
and denoted 
\be
  \mE^2 \equiv \biggl( \frac{\Nf}{6} + \frac{\CA}{3} \biggr)\, g^2 T^2  
  \;. \la{mE}
\ee 
Therefore the divergence can be removed by the usual resummation
of colour-electric modes,  
\be
 - \frac{g^2 C_F \beta}{2} \int_{\vec{k}} \frac{1}{k^2}
 + \frac{g^2 C_F \beta}{2} \int_{\vec{k}} \frac{\mE^2}{k^4} + \ldots 
 \; = \; 
  - \frac{g^2 C_F \beta}{2} \int_{\vec{k}} \frac{1}{k^2+ \mE^2}
 \; = \;
    \frac{g^2 C_F \mE \beta }{8\pi}
 \;. \la{psi_P_LO}
\ee

Unfortunately, 
an inspection of the last row 
in \eq\nr{psi_P_expl} shows that there are also 
logarithmic infrared divergences (cf.\ ref.~\cite{gj}), and 
to handle them correctly we need to proceed carefully. 
A systematic way is through \eq\nr{resum}. 
Within EQCD, the effective Lagrangian has the form 
\be
 {\mathcal{L}}_\rmii{E}  =  
 \fr12 \tr [\tilde{F}_{ij}^2 ]+ \tr [\tilde{D}_i,\tilde{A}_0]^2 + 
 \mE^2\tr [\tilde{A}_0^2] 
 + \ldots 
 \;, 
 \hspace*{0.5cm} \la{eqcd}
\ee
where 
$\tilde{F}_{ij} = (i/\gE) [\tilde{D}_i,\tilde{D}_j]$, 
$\tilde{D}_i = \partial_i - i \gE \tilde{A}_i$, 
$\tilde{A}_i = \tilde{A}^a_i T^a$, 
$\tilde{A}_0 = \tilde{A}^a_0 T^a$, 
and $T^a$ are hermitean generators of SU(3).\footnote{%
 A 2-loop derivation of $\mE^2$, $\gE^2$ in terms 
 of the parameters of four-dimensional QCD can be found in ref.~\cite{gE2}, 
 but here we only need their values at leading non-trivial order,
 $\mE^2 = g^2 T^2 (\Nf/6 + \Nc/3 )$, $\gE^2 = g^2$.}
The Polyakov loop operator is represented as 
\be
 P_\vec{r} = 
 \Bigl[ \unit \, \mathcal{Z}_0 \Bigr]
 + i g \tilde{A}_0 \beta \, \mathcal{Z}_1 
 + \fr12 ( i g \tilde{A}_0 \beta)^2 \, \mathcal{Z}_2 
 + \ldots 
 +  (g^2 \tilde{F}_{ij} \beta^2)^2  \mathcal{X}_4
 + \ldots 
 \;. \la{Pol_EQCD}
\ee
Here all possible local operators made of 
Matsubara zero-modes, invariant under parity and spatial rotations 
and transforming under the adjoint representation of the gauge 
group, can in principle appear. It will be convenient for our 
purposes to use a ``mixed'' convention in \eq\nr{Pol_EQCD}
where $g$ denotes the 
(renormalized) four-dimensional coupling while the fields 
are those of three-dimensional EQCD. The matching 
coefficients $\mathcal{Z}_i$ are of the form 
$\mathcal{Z}_i = 1 + \rmO(g^2)$. The possible 
appearance of a matching coefficient like  
$\mathcal{X}_4$ was pointed out in ref.~\cite{bn2}, but it does
not contribute at the order of our computation. 
The first term in \eq\nr{Pol_EQCD} 
has been put inside brackets, 
because in the language of \eq\nr{resum} 
it represents the value of the unresummed difference
inside the first brackets of \eq\nr{resum}. 

Now, within EQCD, all dynamical effects involve the scale $\mE \sim gT$
and thus bring in additional powers of the coupling $g$. In fact, the 
leading term, originating from the 2nd order operator in \eq\nr{Pol_EQCD}, 
precisely reproduces the result of \eq\nr{psi_P_LO} which is 
of  order $\rmO(g^3)$. Because the term is 
of $\rmO(g^3)$, the coefficient $\mathcal{Z}_2$ can be set 
to unity as we work at the order $\rmO(g^4)$. There is 
an $\rmO(g^4)$ contribution from the 
next-to-leading order evaluation of the 
2nd order operator, however, and the sum of the 
$\rmO(g^3)$ and $\rmO(g^4)$ contributions can be written as  
\ba
 \Bigl[ (\psi_\rmii{P})_\rmii{EQCD} \Bigr]_\rmii{resummed} 
 & = & 
 \frac{g^2 C_F \mE \beta }{8\pi} 
  + \frac{g^2 C_F \beta}{2} \int_{\vec{k}} \! 
     \frac{\Pi_{00}^\rmii{E}(k)}{(k^2 + \mE^2)^2} + \ldots 
 \;, \la{psi_P_E}
\ea 
where the self-energy within EQCD has  a well-known 
form (see, e.g.,\ refs.~\cite{sn1,ar,bn1}):
\ba
 \Pi_{00}^\rmii{E}({k}) & = &  
 \gE^2 \CA T \int_{\vec{q}}
 \biggl\{
   \frac{1}{q^2 + \mE^2} + \frac{2(\mE^2-k^2)}{q^2[(k+q)^2 + \mE^2]} 
   + \nn[2mm] 
 & & \; + \frac{D-3}{q^2} +  \frac{\tilde \xi ( k^2 + \mE^2) }{q^4}
     \biggl[ 1 - \frac{k^2 + \mE^2}{(k+q)^2 + \mE^2} \biggr]
 \biggr\} \la{Pi_E}
 \;. 
\ea
We have kept a general gauge parameter $\tilde\xi$ here, 
defined with the convention that $\tilde\xi = -1$ corresponds 
to Coulomb gauge and $\tilde\xi = 0$ to Feynman gauge
(i.e.\ $G_{ij} = \delta_{ij}/q^2 + \tilde\xi\, q_i q_j/q^4$); 
and, for later reference, 
we have not yet killed any integrals through special properties 
of dimensional regularization. 

We note, first of all, that the $\tilde\xi$-dependent part
of \eq\nr{Pi_E} gives no contribution in \eq\nr{psi_P_E}. 
Second, the other terms of \eq\nr{Pi_E}
reproduce, for $\mE\to 0$, the zero-mode 
contribution on the last line of \eq\nr{psi_P_expl}.
In the ``dressed'' form of \eqs\nr{psi_P_E}, \nr{Pi_E}, however, 
the logarithmic infrared divergence has been lifted. 
The integrals in \eq\nr{psi_P_E} are all elementary
(cf.\ \eqs\nr{ie1}--\nr{ie3}), and in total we get 
(replacing $\gE^2 \to g^2 + \rmO(g^4)$)
\be
 \Bigl[ (\psi_\rmii{P})_\rmii{EQCD} \Bigr]_\rmii{resummed} 
  =  
 \frac{g^2 C_F \mE \beta }{8\pi} 
 - \frac{g^4 C_F \CA}{(4\pi)^2}
 \biggl( \frac{1}{4\epsilon} + 
 \ln\frac{\bmu}{2\mE} + \fr14 \biggr)
 + \rmO(g^5)
 \;. \la{psi_P_E_final}
\ee

%
\subsection{Hard-mode contribution}

It remains to compute the unresummed difference
inside the first brackets in \eq\nr{resum}, given by \eq\nr{psi_P_expl}. 
Without resummation the zero-mode contribution in 
\eq\nr{psi_P_expl} contains no scale and vanishes; 
the same happens in the cases where the $\vec{k}$-integration 
contains no $q_n^2$ and factorizes from the $\vec{q}$-integration. 
This leaves us with 5 non-zero sum-integrals, with
values given in \eqs\nr{P_si1}--\nr{P_si5};
summing together, we arrive at 
\ba
 \mathcal{Z}_0 
 & = &  
 \Bigl[ 
          (\psi_\rmii{P})_\rmii{QCD} - 
          (\psi_\rmii{P})_\rmii{EQCD} \Bigr]_\rmii{unresummed} 
 \nn[3mm] 
 & = & 
   1 + \frac{g^4 C_F}{(4\pi)^2}
   \biggl[
     \Nf \biggl( - \frac{\ln 2}{2} \biggr) 
    + 
     \CA \biggl( \frac{1}{4\epsilon} + \ln\frac{\bmu}{2T} + \fr12
     \biggr)
   \biggr]           
 \;. \la{psi_P_Q_final}
\ea

%
\subsection{Summary and comparison with literature}

Adding up \eqs\nr{psi_P_E_final}, \nr{psi_P_Q_final}, 
the $1/\epsilon$'s duly cancel, and we get our final result for 
the Polyakov loop expectation value: 
\be
 \psi_\rmii{P} = 1 + \frac{g^2 C_F \mE}{8\pi T}
 + \frac{g^4 C_F}{(4\pi)^2}
   \biggl[
     \Nf \biggl( - \frac{\ln 2}{2} \biggr) 
    + 
     \CA \biggl( \ln\frac{\mE}{T} + \fr14
     \biggr)
   \biggr]
 + \rmO(g^5)
 \;, \la{psi_P_final}
\ee
where $\mE$ is from \eq\nr{mE}.\footnote{%
 If the quarks are given a common chemical potential, $\mu$, 
 then the Debye mass parameter gets changed as 
  $
   \mE^2 \to g^2 [\Nc \frac{T^2}{3} + 
  \Nf ( \frac{T^2}{6} + \frac{\mu^2}{2\pi^2} )  ] 
  $, 
 and the numerical factor in \eq\nr{psi_P_final} is modified as 
 $ 
  -\ln 2 / 2 \to \re \zeta'(0, \fr12 + i \frac{\mu}{2\pi T})
 $, where 
 $\zeta'(x,y) \equiv \partial_x \zeta(x,y)$, and 
 $\zeta(x,y)$ denotes the generalized zeta function, 
 $\zeta(x,y)\equiv \sum_{k=0}^\infty 1/(k+y)^x$.     
 }
In accordance with general expectations~\cite{Polyakov:1980ca,smooth}, 
the result is finite and 
renormalization group invariant up to the order computed.\footnote{%
 We have been informed by the authors of ref.~\cite{average} 
 that they have recently obtained the same result. 
 } 

A classic determination of $\psi_\rmii{P}$ was presented, for $\Nf = 0$, 
in ref.~\cite{gj}. Re-expressing that result in terms of $\mE$ 
it can be written as 
\be
 \psi_\rmii{P}^\rmii{\cite{gj}} = 
 1 + \frac{g^2 C_F \mE}{8\pi T}
 + \frac{g^4 C_F \CA}{(4\pi)^2}
     \biggl( \ln\frac{\mE}{2 T} + \fr34
     \biggr)
 \;. \la{psi_P_gj}
\ee
This agrees with our result including the logarithmic term,  
but differs on the constant term accompanying the logarithm 
by a factor $-\ln2 + 1/2$. As far as we can see, the difference
can be traced back to the way that the resummation was carried out.  
We do believe ours to be a systematic resummation.

In order to finally compare with lattice results, 
we need to insert some numerical values for the 
parameters appearing in \eq\nr{psi_P_final}.
Following ref.~\cite{adjoint}, we can estimate them
through some ``fastest apparent convergence'' criteria; 
for the gauge coupling we have applied this criterion to the 
combination $g^2 \mathcal{Z}_1^2$, playing a role in
the next section (cf.\ \eqs\nr{Z12}, \nr{psi_C_final}), 
while for the mass parameter we take over the 
criterion from EQCD~\cite{adjoint}: 
\be
 g^2 \simeq \frac{24\pi^2}{(11\Nc - 2 \Nf)
 [ \ln  \frac{4\pi T }{ \Lambdamsbar }  -\gammaE + c_g ] } 
 \;, \quad 
 \mE^2 \simeq \frac{4\pi^2 (2 \Nc + \Nf) T^2}{(11\Nc - 2 \Nf)
 [ \ln  \frac{4\pi T }{ \Lambdamsbar }  -\gammaE + c_m ] } 
 \;, \quad 
 \la{eff_coup}
\ee
where 
\be
 c_g = \frac{2\Nf(4\ln2 - 1) - 11\Nc}{2(11\Nc - 2 \Nf)}
 \;, \quad
 c_m = \frac{4\Nf \ln2}{11\Nc - 2 \Nf} - 
 \frac{5 \Nc^2 + \Nf^2 + 9 \Nf/(2 \Nc)}{(11\Nc - 2 \Nf)(2 \Nc + \Nf)}
 \;. \la{cs}
\ee
The results
are shown in \fig\ref{fig:psiP}, where they are also compared with 
the four-dimensional lattice data from ref.~\cite{Gupta:2007ax}. 
We have also tested more elaborate choices for the parameters, leading
indeed to a somewhat better accord with lattice data
(cf.\ caption of \fig\ref{fig:psiP}), however within 
the accuracy of our actual computation it is not 
possible to justify these theoretically. Nevertheless experience from other 
quantities, such as the spatial string tension~\cite{mzw,gE2,rbc,hki},  
leads us to suspect that pursuing the computation systematically
to a higher order would eventually allow to improve on the agreement.
(Phenomenological recipes for matching the lattice data down to lower
temperatures can be found, e.g., in refs.~\cite{mrs}.)

It is amusing to note, in any case, that the behaviour of 
the Polyakov loop is qualitatively quite similar to that of
mesonic screening masses, expressed in units
of the temperature~\cite{meson}: both are small close
to the phase transition (because they are related to order
parameters in various limits), but then increase rapidly, and should 
finally approach their non-zero asymptotic values {\em from above}. 

\begin{figure}[t]


\centerline{%
\epsfysize=8.0cm\epsfbox{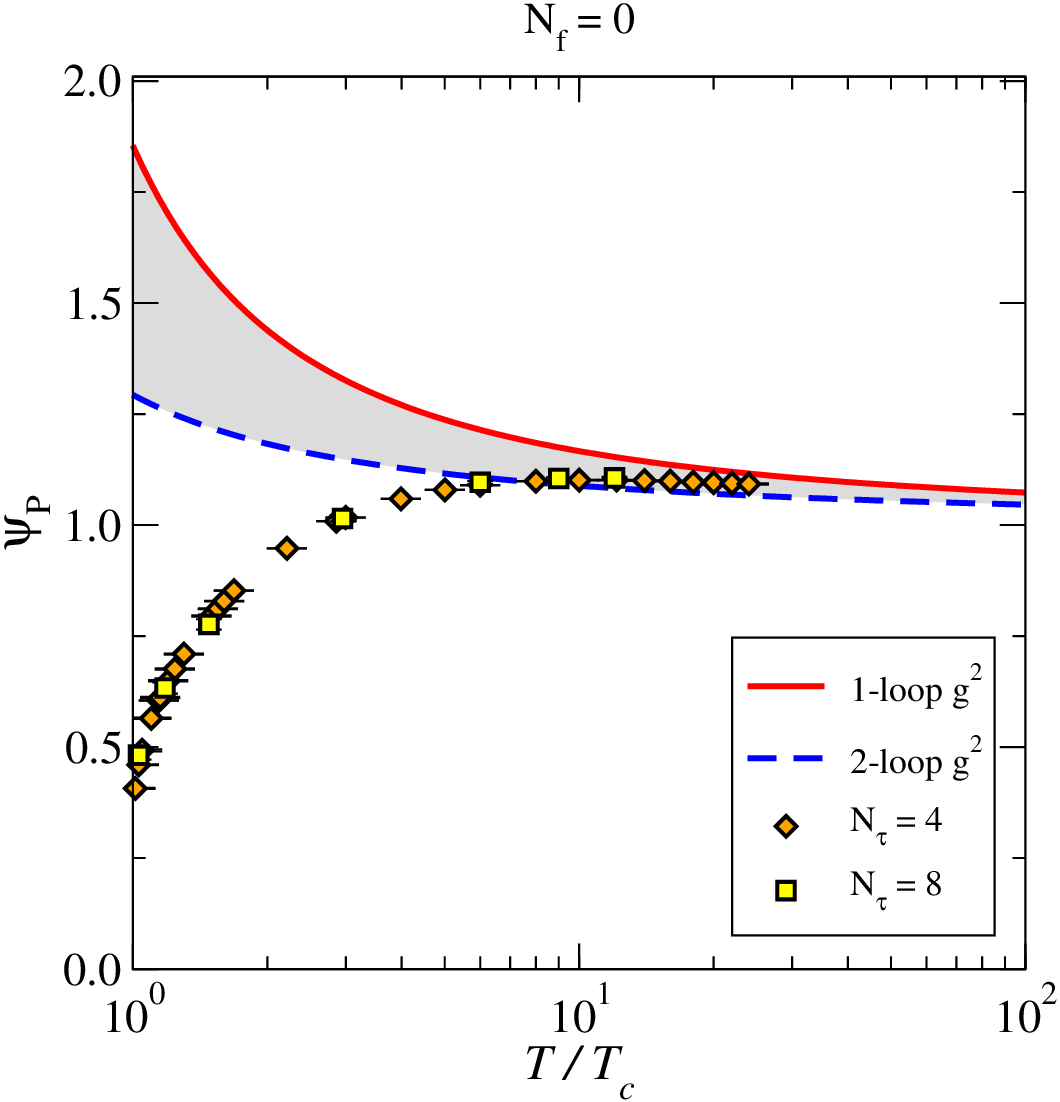}%
}


\caption[a]{\small
The dimensionally regularized Polyakov loop expectation value, 
\eq\nr{psi_P_final}, as a function of $T/\Tc$, where $\Tc$ is
the critical temperature of the deconfining phase transition 
(a conversion from perturbative units has been carried out by
assuming $\Tc/\Lambdamsbar \simeq 1.25$; a variation within 
the range 1.10 -- 1.35 yields an error much smaller than the band width).
The upper edge of the band (solid red line) 
corresponds to evaluating the coupling
and the Debye mass parameter according to the simple 1-loop criteria
in \eqs\nr{eff_coup}, \nr{cs}; for the lower edge 
(dashed blue line) we have 
replaced $g^2$ through the 2-loop value of $\gE^2$ given 
in ref.~\cite{gE2}, and $\mE/\gE^2$ through the expression
in eq.~(14) of ref.~\cite{lv}. The lattice data, 
labelled by $N_\tau$, is from ref.~\cite{Gupta:2007ax}
(the spatial lattice size was kept fixed at $32^3$).
}

\la{fig:psiP}
\end{figure}

%
\section{Singlet free energy in Coulomb gauge}
\la{se:Coulomb}

%
\subsection{Basic setup}

The original definition of the singlet quark-antiquark free energy
was related to the eigenvalues of the untraced Polyakov loop~\cite{sn2};  
in practice, however, lattice measurements consider 
the object (see, e.g., \cite{q_singlet}--\cite{whot} and references therein) 
\be
 \psi_\rmii{C}(r) \equiv
 \frac{1}{\Nc}
 \langle \tr[P_\vec{r} P^\dagger_\vec{0} ] \rangle_\rmi{Coulomb}
 \;. \la{Vsinglet}
\ee
This differs from the physical ``colour-averaged'' 
free energy, in which traces of Polyakov loops are 
correlated (cf.\ \eq\nr{psi_T_def})~\cite{mls,sn1,jp}; 
the advantage of \eq\nr{Vsinglet}
is that the free energy extracted from it is believed 
to reduce at small distances to the gauge-fixing independent 
zero-temperature static potential~\cite{fz,whot}.

%
\begin{figure}[t]
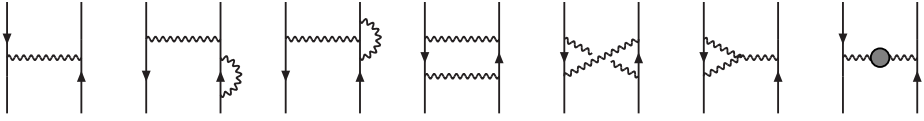


\begin{eqnarray*}
&& 
 \hspace*{-2cm}
 \CoulA 
 \CoulB 
 \CoulC 
 \CoulD 
 \CoulE 
 \CoulF 
 \CoulG 
\end{eqnarray*}

\caption[a]{\small 
The connected graphs contributing to the correlator of two Polyakov 
loops at $\rmO(g^4)$, with the filled blob denoting the 1-loop
gauge field self-energy. In addition there are disconnected 
contributions, obtained by multiplying the set of graphs in \fig\ref{fig:Poly} 
with its Hermitean conjugate. (The topologies are the same as those needed
for pushing the computation of the heavy quark medium polarization one 
level up from the current standard~\cite{nlo}.)} 
\la{fig:Coulomb}
\end{figure}
%

The general strategy for our determination of $\psi_\rmii{C}$ is the same 
as in \se\ref{se:pol}. The new graphs to be evaluated 
are shown in \fig\ref{fig:Coulomb}. Like in the previous
section, we first give a general (unresummed) result, 
which is valid in any gauge with the property $G_{0i}(0,\vec{k}) = 0$:
\ba
 \psi_\rmii{C}(r) & = & 
 \exp\biggl[ 
   g^2 C_F \beta \int_{\vec{k}} \! 
   \Bigl(e^{i \vec{k}\cdot\vec{r}} -1 \Bigr)
   G_{00}(0,\vec{k})
 \biggr] 
 \nn & & \; 
 - g^2 C_F \beta \int_{\vec{k}} \! 
   \Bigl(e^{i \vec{k}\cdot\vec{r}} -1 \Bigr)
   G_{00}^2(0,\vec{k}) \, \Pi_{00}(0,\vec{k})
 \nn & & \; 
  + g^4 C_F \CA  
   \int_{\vec{k}} \! 
   \Bigl(e^{i \vec{k}\cdot\vec{r}} -1 \Bigr)
   G_{00}(0,\vec{k}) \int_{\vec{q}}
        \biggl\{ 
        \frac{\beta^2 G_{00}(0,\vec{q})}{6} - 
        \sum_{q_n\neq 0} \frac{2 G_{00}(q_n,\vec{q})}{q_n^2}
   \nn & & \; \hspace*{1.5cm}
   - 2  \sum_{q_n\neq 0} 
     \biggl[
        G_{00}(q_n,\vec{q}) \frac{k_i - q_i}{q_n} +  
        G_{0i}(q_n,\vec{q})
     \biggr] G_{0i}(q_n,\vec{q+k})    \biggr\} 
   \nn & & \; 
  + g^4 C_F \CA  
   \int_{\vec{k},\vec{q}} \! 
   \Bigl(e^{i (\vec{k+q})\cdot\vec{r}} -1 \Bigr)
   \biggl\{
     -\frac{\beta^2 G_{00}(0,\vec{q}) G_{00}(0,\vec{k})}{8}
   \nn & & \; \hspace*{1.5cm}
     + \sum_{q_n\neq 0} \frac{G_{00}(q_n,\vec{q})}{q_n^2}
     \Bigl[ G_{00}(0,\vec{k}) + \fr12 G_{00}(q_n,\vec{k})
     \biggr] 
   \biggr\}
 \;. \la{general_r} \la{psi_C_full}
\ea
This was obtained by a brute force evaluation, 
making use of \eqs\nr{tri1}, \nr{tri2}, etc.
The result is gauge-dependent; 
in order to now specialize to Coulomb gauge, we insert the gluon propagator
\be
 G_{\mu\nu}(Q) = 
 \frac{ \delta_{\mu 0} \delta_{\nu 0} }{\vec{q}^2}
 + 
 \frac{ \delta_{\mu i} \delta_{\nu j} }{Q^2} 
 \Bigl( 
   \delta_{ij} - \frac{q_i q_j}{\vec{q}^2}
 \Bigr) 
 \;, \quad Q=(q_n,\vec{q})
 \;, \la{G_C} 
\ee
and the corresponding self-energy into \eq\nr{general_r}. 
The self-energy reads 
\ba
 \Pi_{00}(0,\vec{k}) & = & 
 g^2 \CA 
 \Tint{Q}
 \biggl\{ 
   \frac{D-2}{Q^2} -  \frac{2[k^2 + (D-2) q_n^2]}{Q^2(Q+K)^2}
    \nn & & \;  
   - \frac{k^2}{q^2 Q^2}
   + \frac{k^4}{q^2(q+k)^2Q^2} 
   - \frac{k^4 q_n^2}{2 q^2 (q+k)^2 Q^2(Q+K)^2}
 \biggr\}
   \nn &  + &  
 g^2 \Nf \Tint{\{ Q \} }
 \biggl\{ 
   -\frac{2}{Q^2} + \frac{k^2 + 4 q_n^2}{Q^2(Q+K)^2}
 \biggr\} 
 \;, \la{Pi_C}
\ea
where again $K\equiv (0,\vec{k})$. 
To be explicit, the result can be written as
\ba
 \psi_\rmii{C}(r) & = & 
 1 + {g^2 C_F \beta} 
 \int_{\vec{k}} \frac{e^{i \vec{k}\cdot\vec{r}}-1}{k^2}
 + \fr12 \biggl( g^2 C_F \beta \int_{\vec{k}}
 \frac{e^{i \vec{k}\cdot\vec{r}}-1}{k^2} 
   \biggr)^2
 \nn & & \; 
 + g^4 C_F \Nf \sum_{ \{ q_n \} } \int_{\vec{k,q}}
 \Bigl( e^{i \vec{k}\cdot\vec{r}}-1 \Bigr)
 \biggl[ 
 \frac{2}{k^4 Q^2} 
 - \frac{1}{k^2 Q^2 (Q+K)^2}
 - \frac{4 q_n^2}{k^4 Q^2(Q+K)^2}
 \biggr]
 \nn & & \; 
  + g^4 C_F \CA  \sum_{ q_n } \int_{\vec{k,q}}
 \Bigl( e^{i \vec{k}\cdot\vec{r}}-1 \Bigr)
 \biggl[ 
   \frac{2-D}{k^4 Q^2}
 + \frac{2 }{k^2 Q^2 (Q+K)^2}
 + \frac{2 (D-2) q_n^2}{k^4 Q^2(Q+K)^2}
 \nn & & \; \hspace*{3cm}
 + \frac{1}{k^2 q^2 Q^2}
 - \frac{1}{2 q^2 Q^2 (q+k)^2}
 - \frac{1}{2 q^2 Q^2 (Q+K)^2}
 \biggr]
 \nn & & \; 
 + {g^4 C_F \CA} \int_{\vec{k}} 
  \frac{e^{i \vec{k}\cdot\vec{r}}-1}{k^2} 
 \biggl[
   \biggl( \frac{\beta^2}{6} - \sum_{q_n'} \frac{2}{q_n^2} \biggr)
   \int_{\vec{q}} \frac{1}{q^2}
 \biggr]
 \nn & & \; 
 + {g^4 C_F \CA} \int_{\vec{k,q}} 
  \frac{e^{i (\vec{k+q})\cdot\vec{r}}-1}{k^2} 
 \biggl[
   \biggl( - \frac{\beta^2}{8} + \sum_{q_n'} \frac{3}{2 q_n^2} \biggr)
   \frac{1}{q^2}
 \biggr]
 \;. \la{psi_C_expl}
\ea
Given that 
$
 \sum_{q_n'} q_n^{-2} = \beta^2 \zeta(2) / 2 \pi^2 = \beta^2/12
$, 
the last two terms (corresponding to a vertex correction and to two-gluon 
exchange) actually do not contribute; in Coulomb
gauge the unresummed result originates from 
the self-energy correction alone.   

If we naively set $|\vec{r}|\to\infty$ in \eq\nr{psi_C_expl}, 
then all terms containing $e^{i \vec{k}\cdot\vec{r}}$ or 
$e^{i (\vec{k+q})\cdot\vec{r}}$ drop out. A constant term
remains over, and as can be verified by comparing with  
\eq\nr{psi_P_expl}, 
it has the value\footnote{%
 The expression in \eq\nr{psi_P_expl} had the covariant gauge 
 as a starting point whereas \eq\nr{psi_C_expl} comes from
 the Coulomb gauge; therefore some of the terms do not immediately
 look alike. On closer inspection, however, their difference 
 integrates to zero in the $\vec{r}$-independent term.
 } 
\be
 \lim_{r \to \infty} \psi_\rmii{C}(r) = |\psi_\rmii{P}|^2  
 \;.
\ee
In the following we drop out this constant contribution, 
and focus on the $r$-dependent terms. 

%
\subsection{Short-distance limit} 
\la{psi_C_short}

The Matsubara sums in \eq\nr{psi_C_expl} can be transformed, 
in a standard way, to an integral representation, from which
a zero-temperature part and a thermal part can be identified. 
The thermal part contains the scale $T$ inside Bose-Einstein
and Fermi-Dirac distribution functions and, for large momenta, 
is exponentially suppressed. Therefore we expect that at small
distances, i.e.\ $r \ll \frac{1}{\pi T}$, the result can be 
obtained by simply replacing the Matsubara sums with the 
corresponding zero-temperature integrals, 
$
 \Tinti{Q}, \Tinti{ \{ Q \} } 
 \to \int_Q 
$.
(Of course, this expectation can be crosschecked later on from
the more general result.)

The advantage of the zero-temperature limit is that then 
all integrals can be carried out analytically. For the 
``covariant'' structures in \eq\nr{psi_C_expl} we get, as usual, 
\ba
 \int_Q \frac{q_0^2}{Q^2(Q+K)^2} & = & 
 \frac{1}{D-1} \int_Q
 \biggl[ 
   \frac{1}{2 Q^2} - \frac{k^2}{4 Q^2(Q+K)^2}
 \biggr]
 \;, \\
 \int_Q \frac{1}{Q^2} & = & 0 
 \;, \\ 
 \int_Q \frac{1}{Q^2(Q+K)^2} & = &  
 \frac{1}{(4\pi)^2}
 \biggl( \frac{1}{\epsilon} + \ln\frac{\bmu^2}{k^2} + 2 \biggr)
 \;,   
\ea
where we once again inserted $K=(0,\vec{k})$.
As far as the ``non-covariant'' terms are concerned, 
it is convenient to group them as 
\ba
 & & 
 \int_Q \frac{1}{ q^2 Q^2} 
 \biggl[ \frac{1}{k^2}
 - \frac{1}{2 (q+k)^2}
 - \frac{1}{2 (Q+K)^2} \biggr]
 \nn & = & 
 \int_Q \frac{1}{q^2 Q^2}
 \biggl[ \frac{1}{k^2} - \frac{1}{(q+k)^2} \biggr]
 + 
 \int_Q \frac{1}{2 q^2 Q^2}
 \biggl[ \frac{1}{(q+k)^2}
 - \frac{1}{(Q+K)^2} \biggr]
 \nn & = &  
 \frac{1}{8\pi^2 k^2} \biggl( \frac{1}{\epsilon} + 
 \ln\frac{\bmu^2}{4 k^2} \biggr) 
 \hspace*{0.75cm} + 
 \frac{1}{8\pi^2 k^2} \ln 4
 \;. 
\ea
The former combination is infrared finite but ultraviolet divergent, 
and the integral was carried out in dimensional regularization; 
the latter combination is infrared and ultraviolet finite 
for $\vec{k}\neq 0$, and was integrated directly in four dimensions. 

Summing up and re-expanding the bare gauge coupling of the leading 
order term in terms of the renormalized one, the $1/\epsilon$'s
cancel (more details on the graph-by-graph origin of divergences 
are provided in \se\ref{se:xi} for a case where they do {\em not} cancel), 
and we get
\ba
 \psi_\rmii{C}(r) & \stackrel{r \pi T \ll 1}{\approx} & 
 \exp\biggl( \frac{g^2 C_F \beta}{4\pi r} \biggr)
 \nn & + & 
 \frac{g^4 C_F \beta}{(4\pi)^2}
 \int_{\vec{k}} \frac{e^{i \vec{k}\cdot\vec{r}}}{k^2}
 \biggl[
  -\frac{2\Nf}{3} 
  \biggl( 
    \ln\frac{\bmu^2}{k^2} + \fr53
  \biggr) 
  + \frac{11 \CA}{3}
  \biggl( 
    \ln\frac{\bmu^2}{k^2} + \fr{31}{33}
  \biggr) 
 \biggr]
 \;. \la{zeroT}
\ea
Exponentiating the $\rmO(g^4)$ correction and extracting the singlet 
free energy, $V_1$, as
$
 \psi_\rmii{C}(r) \equiv \exp[-\beta V_1(r)]
$, 
we see that $V_1$ precisely agrees with the classic result for 
the zero-temperature static potential at $\rmO(g^4)$~\cite{wf}\footnote{%
 We recall that in the Fourier transform the logarithm 
 gets effectively replaced as 
 $ 
  \ln(\bmu^2/k^2) \to 2 [\ln(\bmu r) + \gammaE]
 $. \la{f4}
 }:
\be
 V_1(r) = - \frac{g^2 C_F}{4\pi r}
 + \frac{g^4 C_F}{(4\pi)^2}
 \int_{\vec{k}} \frac{e^{i \vec{k}\cdot\vec{r}}}{k^2}
 \biggl[
  \frac{2\Nf}{3} 
  \biggl( 
    \ln\frac{\bmu^2}{k^2} + \fr53
  \biggr) 
  - \frac{11 \CA}{3}
  \biggl( 
    \ln\frac{\bmu^2}{k^2} + \fr{31}{33}
  \biggr) 
 \biggr]
 \;.
\ee
Empirical observations in the same direction have been 
made even on the non-perturbative level~\cite{fz,whot}; thus the 
agreement might be true at the 2-loop level as well, where 
a comparison could be carried out with ref.~\cite{ys}, 
however we have not undertaken this task.  

%
\subsection{Hard-mode contribution}

We now move on to consider the behaviour of \eq\nr{psi_C_expl}
at larger distances, $r\sim 1/\pi T$, 
where thermal effects do play a role. The general
strategy is analogous to \eq\nr{resum}, {\it viz.}\
\be
 \psi_\rmii{C}(r) = \Bigl[ 
          (\psi_\rmii{C}(r))_\rmii{QCD} - 
          (\psi_\rmii{C}(r))_\rmii{EQCD} \Bigr]_\rmii{unresummed} + 
          \Bigl[ (\psi_\rmii{C}(r))_\rmii{EQCD} \Bigr]_\rmii{resummed} 
 \;. \la{resum2}
\ee
This time we evaluate the unresummed QCD part first, 
and for ease of later subtraction separate
the Matsubara zero mode contribution from \eq\nr{psi_C_expl}, 
i.e.\ replace $\sum_{q_n} \to \sum_{q_n'}$, and write 
the zero-mode part separately.

For the non-zero mode part, we proceed as follows. The goal is to carry 
out the sum-integral over $Q$ of non-zero Matsubara frequency, 
and to express the resulting 
$\vec{k}$-integrand, let us call it $\mathcal{I}(k)$, in the form 
\be
 \mathcal{I}(k) = \frac{\mathcal{A}}{k^4} + \frac{\mathcal{B}}{k^2} 
 + \mathcal{C}(k)
 \;. \la{strategy}
\ee
Because of restriction to non-zero modes, the function $\mathcal{C}(k)$
must be analytic in $k^2$. 
Like in \eq\nr{psi_P_LO}, the coefficient $\mathcal{A}$ is 
essentially $\mE^2$, and this most infrared sensitive term is 
subtracted by the expanded version of the 
EQCD contribution. The coefficient $\mathcal{B}$, in turn, 
``renormalizes'' the leading order contribution; in fact 
it also gets subtracted by the EQCD contribution
through \eq\nr{resum2}, more precisely 
by the 1st order term from \eq\nr{Pol_EQCD}, 
with a properly chosen factor $\mathcal{Z}_1$. 
The remainder, determined by the 
function $\mathcal{C}(k)$, represents the ``genuine'' 
thermal contribution from the non-zero Matsubara modes, 
and does not get subtracted by EQCD effects.

Making use of the sum-integrals in 
\eqs\nr{si1}, \nr{si2}, \nr{si6}, \nr{rint_1}--\nr{rint_6}, 
we get 
\ba
 & & \hspace*{-1cm} 
 \Bigl[(\psi_\rmii{C}(r))_\rmi{QCD}\Bigr]_\rmi{unresummed}
 \nn  \!\! & = & \!\! 
 \mbox{const.} + {g^2 C_F \beta} 
 \int_{\vec{k}} \frac{e^{i \vec{k}\cdot\vec{r}}}{k^2}
 + \fr12 \biggl( g^2 C_F \beta \int_{\vec{k}}
 \frac{e^{i \vec{k}\cdot\vec{r}}}{k^2} 
   \biggr)^2
 \nn & & \; 
 + g^2 C_F \beta \int_{\vec{k}} e^{i \vec{k}\cdot\vec{r}} 
 \biggl\{ 
   -\frac{\mE^2}{k^4}
   +\frac{g^2}{(4\pi)^2 k^2}
   \biggl(
     \frac{11\Nc}{3} ( 
                              L_\bo + 1 )  
    - 
     \frac{2\Nf}{3} 
   (
      L_\fe - 1 
   )  
   \biggr)
 \biggr\}  
 \nn & & \; 
  + \frac{g^4 C_F \CA}{(4\pi)^2} 
 \biggl[ 
  - \frac{1}{24 T^2 r^2}
  + \frac{5 \ln(1-e^{-4\pi T r})}{6\pi T r}
  + \frac{1 + e^{4\pi T r}(4\pi T r - 1)}
         {3(e^{4\pi T r}-1)^2}
 \nn & & \; \hspace*{3cm}
 + \sum_{n=1}^{\infty} \biggl( 4 -
   \frac{    (4\pi T r n)^2 }{3}  \biggr)
    E_1\Bigl(4\pi T r n\Bigr)
 + \frac{{\rm Li}_2(e^{-4\pi T r})}{(2\pi T r)^2}
 \biggr]
 \nn & & \; 
 + \frac{g^4 C_F \Nf}{(4\pi)^2} 
 \biggl[ 
  - \frac{1}{3\pi T r}  \ln\frac{1-e^{-2\pi T r}}{1+e^{-2\pi T r}}
   + \frac{e^{2\pi T r}}{3}
    \biggl(
      \frac{1}{e^{4\pi T r}-1} - 2 \pi T r 
      \frac{e^{4\pi T r}+1}{(e^{4\pi T r}-1)^2}
    \biggr)
 \nn & & \; \hspace*{3cm}
 + \sum_{n=1}^{\infty} \biggl( -2  + 
   \frac{    [2\pi T r \times (2 n - 1)]^2 }{3}  \biggr)
    E_1\Bigl(2\pi T r  \times (2 n - 1) \Bigr)
 \biggr]
 \nn & & \; 
 + g^4 C_F \CA \times \mbox{(zero mode contribution)}
 \;. \la{psi_C_inter}
\ea
Here the gauge coupling  is already the renormalized one;
$L_\bo, L_\fe$ are defined as
\be
 L_\bo \equiv 2 \ln \frac{\bmu e^{\gammaE}}{4\pi T}
  \;, \quad
 L_\fe \equiv 2 \ln \frac{\bmu e^{\gammaE}}{\pi T}
  \;;   \la{log_defs}
\ee
and $E_1$ and ${\rm Li}_2$ are defined in \eqs\nr{E1}, \nr{Li2}.
Sometimes it may be convenient to replace the sums over $E_1$
with integral representations, and indeed this can be achieved 
as specified in \eqs\nr{int_rep}--\nr{int_rep_last}. 
The expression then compactifies quite a bit; 
we rewrite the corresponding result in \eq\nr{psi_C_final} below.

As a first check, it can be shown that for $\pi T r \ll 1$
the square-bracketed terms in \eq\nr{psi_C_inter} go over 
into 
\be
 \frac{g^4 C_F \CA}{(4\pi)^3Tr} \times
 \frac{11}{3} 
 \biggl[ 2 \ln(4\pi T r) - 1 + \frac{31}{33} \biggr] - 
 \frac{g^4 C_F \Nf}{(4\pi)^3Tr} \times
 \frac{2}{3} 
 \biggl[ 2 \ln(\pi T r) + 1 + \frac{5}{3} \biggr]
 \;.
\ee
Combining with the logarithmic terms 
in \eq\nr{psi_C_inter}, 
we exactly
match the behaviour given by \eq\nr{zeroT}. 

%
\subsection{Soft-mode contribution}

We then proceed to consider the EQCD (zero-mode) 
contribution to \eq\nr{resum2}, which we now denote by 
$\psi_\rmii{C}^\rmii{E}(r) \equiv (\psi_\rmii{C}(r))_\rmii{EQCD}$. 
It reads
\ba
 \psi_\rmii{C}^\rmii{E}(r) & = & \mbox{const.} + 
 g^2 C_F \beta 
 \mathcal{Z}_1^2 \int_{\vec{k}} \frac{e^{i\vec{k}\cdot\vec{r}}}{k^2 + \mE^2}
 + 
 \fr12 \biggl( 
  g^2 C_F \beta  
 \mathcal{Z}_2 \int_{\vec{k}} \frac{e^{i\vec{k}\cdot\vec{r}}}{k^2 + \mE^2}
 \biggr)^2
 \nn & & + \; 
 g^2 \gE^2 C_F \CA \mathcal{Z}_1^2  \int_{\vec{k,q}} e^{i\vec{k}\cdot\vec{r}}
 \biggl\{
       \frac{1}{(k^2 + \mE^2)^2}
    \biggl[ \frac{3-D}{ q^2} 
    -  \frac{1}{q^2 + \mE^2 } \biggr]
   \nn & & \qquad + \; 
      \frac{2}{(k^2 + \mE^2)[(k+q)^2 + \mE^2] q^2} 
    - \frac{4 \mE^2 }{(k^2 + \mE^2)^2 [(k+q)^2 + \mE^2] q^2} 
   \nn & & \qquad + \; 
    \frac{1 }{q^4}
     \biggl[ \frac{1}{k^2 + \mE^2} - \frac{1}{(k+q)^2 + \mE^2} \biggr]
 \biggr\}
 \nn & & - \; 
  \frac{g^4 C_F \CA }{8}
 \biggl( 
 \beta  
 \mathcal{Z}_2 \int_{\vec{k}} \frac{e^{i\vec{k}\cdot\vec{r}}}{k^2 + \mE^2}
 \biggr)^2  + \rmO(g^5)
 \la{psi_C_E}
 \;. 
\ea
In the terms of $\rmO(g^4)$, we can immediately set 
$\gE^2\to g^2$, 
$\mathcal{Z}_1, \mathcal{Z}_2 \to 1$.

Let us inspect the role that \eq\nr{psi_C_E} plays
when first subtracted in an unresummed form, and then added 
``as is'', to \eq\nr{psi_C_inter}. The first line 
of \eq\nr{psi_C_E} accounts for the first two lines 
of \eq\nr{psi_C_inter}, cancelling the infrared divergence 
like in \eq\nr{psi_P_LO} and fixing 
\be
 \mathcal{Z}_1^2 = 1 + \frac{g^2}{(4\pi)^2}
    \biggl[
     \frac{11\Nc}{3} \biggl( 
                              L_\bo + 1 \biggr)  
    - 
     \frac{2\Nf}{3} 
   \biggl(
      L_\fe - 1 
   \biggr)  
   \biggr]
 \;. \la{Z12}
\ee
The expression in the curly brackets in \eq\nr{psi_C_E} agrees, for 
$\mE\to 0$, with the zero-mode part of \eq\nr{psi_C_expl}, 
and through the subtraction--addition step
replaces it with a ``less'' infrared sensitive expression; 
we return to this term presently. Finally, the last 
term of \eq\nr{psi_C_E} can be written, after subtraction
and addition and insertion $\mathcal{Z}_2\to 1$, as 
\be
 \Bigl[ 
           - 
          \psi^\rmii{E}_\rmii{C}(r) \Bigr]_\rmii{unresummed} + 
          \Bigl[ \psi^\rmii{E}_\rmii{C}(r) \Bigr]_\rmii{resummed} 
  = \; \ldots \; + \;
 \frac{g^4 C_F \CA}{(4\pi)^2}
 \biggl[
   \frac{1}{8 T^2 r^2}
   - 
   \frac{\exp(-2 \mE r)}{8 T^2 r^2} 
 \biggr]
 \;. \quad
 \la{powerlaw}
\ee

Now, we observe a problem. After combining \eq\nr{psi_C_inter}
with the re-processed version of \eq\nr{psi_C_E}, 
as outlined above, 
the large-distance behaviour of $\psi_\rmii{C}$ is dominated
by an uncancelled {\em power-law term}, 
\be
 \psi_\rmii{C}(r) \; \stackrel{\pi T r\gg 1}{\approx} \;
  \frac{g^4 C_F \Nc}{(4\pi)^2} 
  \biggl[ -  \frac{1}{24 T^2 r^2} + \frac{1}{8 T^2 r^2}  \biggr]
 = 
  \frac{g^4 C_F \Nc}{(4\pi)^2} 
  \biggl[ \frac{1}{12 T^2 r^2}  \biggr]
 \;. \la{problem} 
\ee
Indeed all other terms are exponentially suppressed after the
subtraction--addition step, 
either as $\exp(-2\pi T r)$ or (from \eq\nr{psi_C_E})
as $\exp(-\mE r)$ or $\exp(-2 \mE r)$. A term of the type
in \eq\nr{problem} must be a gauge artifact: 
physically there is a finite screening length in 
a non-Abelian plasma.\footnote{%
 Within perturbation theory a power-law term could in principle
 also indicate a sensitivity to colour-magnetic modes; 
 however, as \eq\nr{powerlaw} suggests, the term here has 
 at least partly a colour-electric origin. 
 } 
Indeed we will find in \se\ref{se:cyclic}
(cf.\ \eq\nr{psi_W_inter}) that in gauge-invariant observables 
the power-law terms do get duly cancelled. 

Despite the issue of \eq\nr{problem}, which we consider to be 
a serious one from the conceptual point of view, 
we wish to complete in the remainder of this section
our discussion of \eq\nr{psi_C_E}. This leads to another issue, 
yet this time more physical, being a manifestation 
of the logarithmic sensitivity 
of the result to the non-perturbative colour-magnetic scale. 

First of all, if we just carry out the integrals
in the curly brackets of \eq\nr{psi_C_E} literally, 
the result appears to be both
infrared and ultraviolet finite, and reads
\ba
 \psi_\rmii{C}^\rmii{E}(r) \!\! & = & \!\! 
 \mbox{const.}\ + 
  \frac{g^2 C_F \mathcal{Z}_1^2  e^{-\mE r}}{4\pi T r} + 
 \fr12 \biggl( 
 \frac{g^2 C_F   e^{-\mE r}}{4\pi T r} \biggr)^2
 \nn[3mm]  &  &  + \;  
 \frac{g^4 C_F \CA e^{-\mE r}}{(4\pi)^2}
 \biggl\{ 
   2 - \ln(2\mE r) - \gammaE + e^{2 \mE r} E_1 (2 \mE r)
 \biggr\}
 \nn[3mm] & &  -\; \frac{g^4 C_F \CA}{(4\pi)^2}
   \frac{\exp(-2 \mE r)}{8 T^2 r^2} + \rmO(g^5) 
 \la{psi_E_final}
 \\[3mm] & \stackrel{r\gg \pi/g^2 T}{\approx} &  
 \frac{g^4 C_F \CA e^{-\mE r}}{(4\pi)^2}
 \biggl[
   - \ln (\mE r) + 2 - \ln2 - \gammaE
   + \rmO\Bigl(\frac{1}{\mE r}\Bigr) 
 \biggr] + \rmO(g^5)
 \;. \hspace*{5mm}
\ea
On the last row we displayed the large-distance behaviour. 
However, the logarithmic dependence on $r$ 
(see also ref.~\cite{sn1}) implies that this term {\em cannot}
be interpreted as a mass correction to the leading order result, 
but that the correction of $\rmO(g^4)$ overtakes the correction of 
$\rmO(g^2)$ for $r \gg \pi/g^2 T$, rendering the perturbative series out 
of control. A further issue is that the gauge dependent term, the last one
within the curly brackets in \eq\nr{psi_C_E}, does give a finite 
non-zero contribution to \eq\nr{psi_E_final}. 
These well-known issues have lead to attempts 
at interpreting the self-energy insertion in 
a different way~\cite{ar}, and we now discuss these. 

The idea is to treat the 1-loop self-energy 
as if it were an analytic function of $k^2$, as would be the case 
if there were no infrared problems.  In this case, making use of 
the symmetry in $k\to -k$, we can write the self-energy contribution
(curly brackets in \eq\nr{psi_C_E}) as
\ba
 \psi^\rmii{E}_\rmii{C}(r) & = & \ldots  
 + g^2 C_F \beta 
 \int \! \frac{{\rm d}^3\vec{k}}{(2\pi)^3} e^{i\vec{k}\cdot\vec{r}}
 \biggl[ \frac{1}{k^2 + \mE^2} - 
 \frac{\Pi_{00}^\rmii{E}(k)}{(k^2 + \mE^2)^2} \biggr]
 \nn & = & 
 \ldots
 + \frac{g^2 C_F \beta}{4\pi^2 i r}
 \int_{-\infty}^{\infty} \! {\rm d}k \, k \, e^{i k r}
 \biggl[
    \frac{1}{k^2 + \mE^2}
   - \frac{
     \Pi_{00}^\rmii{E}(i \mE) 
     + (k - i \mE ) 
     {\Pi_{00}^\rmii{E}}'(i \mE) 
     + ...
     }{(k^2+ \mE^2)^2} 
 \biggr] 
 \nn & \approx & 
  \ldots 
 + \frac{g^2 C_F }{4\pi T r}
   \biggl[    
     1 - \frac{{\Pi_{00}^\rmii{E}}'(i \mE)}{2 i \mE}
   \biggr]
   \exp\biggl\{
        - \biggl( \mE +  \frac{\Pi_{00}^\rmii{E}(i \mE)}{2 \mE} 
       \biggr) r
       \biggr\} 
 \;, \la{resum_mass}
\ea
where we closed the contour in the upper half plane and 
then resummed the mass correction into the leading order term. 
Inserting the values (cf.\ \eq\nr{Pi_E})
\ba
 \frac{\Pi_{00}^\rmii{E}(i \mE)}{2 \mE} 
 & = & 
 -\frac{g^2 \CA T}{8\pi}
 \biggl( \frac{1}{\epsilon_\rmii{IR}} + 
 \ln\frac{\bmu^2}{4\mE^2}  + 1 
 \biggr)
 \;, \la{PiE} \\  
 \frac{{\Pi_{00}^\rmii{E}}'(i \mE)}{2 i \mE}
 & = &  0 
 \;, 
\ea
we note that handled this way, 
the dependence on the gauge parameter $\tilde\xi$ disappears; 
however, an uncancelled infrared divergence, 
associated with contributions from the colour-magnetic scale, is left 
over. In fact the term in \eq\nr{PiE} equals
the ultraviolet matching coefficient from the 
scale $\mE$ that plays a role in the approach of ref.~\cite{ay} 
(with the only difference that a subtraction of an unresummed infrared
contribution, 
$g^2 \Nc T/8\pi \times 
(1/\epsilon_\rmii{UV} - 1/ \epsilon_\rmii{IR}) = 0$, 
transforms $1/\epsilon_\rmii{IR}$ into $1/\epsilon_\rmii{UV}$
in that case).

The third possibility is to use the pole mass
method but to ``regulate'' the infrared behaviour
of the spatial gluon propagators by introducing a ``magnetic mass''
as a regulator~(see, e.g., refs.~\cite{ar,bn1}). In this way a finite
and gauge-independent result is obtained; however, it is ambiguous, 
because the value of the magnetic mass 
has no well-defined meaning.

On the non-perturbative level, we expect that the infrared problem
is cured by physics at the colour-magnetic scale $g^2 T/\pi$; 
this physics being that of three-dimensional confinement, it however 
cannot be reduced to a simple magnetic mass. Nevertheless, we could
still expect exponential decay as in \eq\nr{resum_mass}, with a 
non-perturbative version of the correction in \eq\nr{PiE}, that is 
with the screening mass
\be
 \tilde{m}_\rmii{E} = \mE + \frac{g^2 \CA T}{4\pi}
 \biggl(\ln\frac{\mE}{g^2 T} + c_\rmii{E} \biggr) + \rmO(g^3 T)
 \;. \la{tmE}
\ee
If the mass $\tilde{m}_\rmii{E}$ were 
dictated by the mass of the lightest gauge-invariant state, 
in the spirit of ref.~\cite{ay}, which
in perturbation theory reduces to the structure of \eq\nr{PiE},  
then $c_\rmii{E} \approx 6.9$ for $\Nc = 3$~\cite{lp2}. This is just a guess, 
however; the result can as well be gauge-dependent, 
given that the large distance $r$-dependence of the Coulomb 
gauge correlator appears in any case to be determined 
by an unphysical power-law contribution, as discussed above. 
Therefore we treat $\tilde{m}_\rmii{E}$ as a free parameter for now. 

\begin{figure}[t]


\centerline{%
\epsfysize=8.0cm\epsfbox{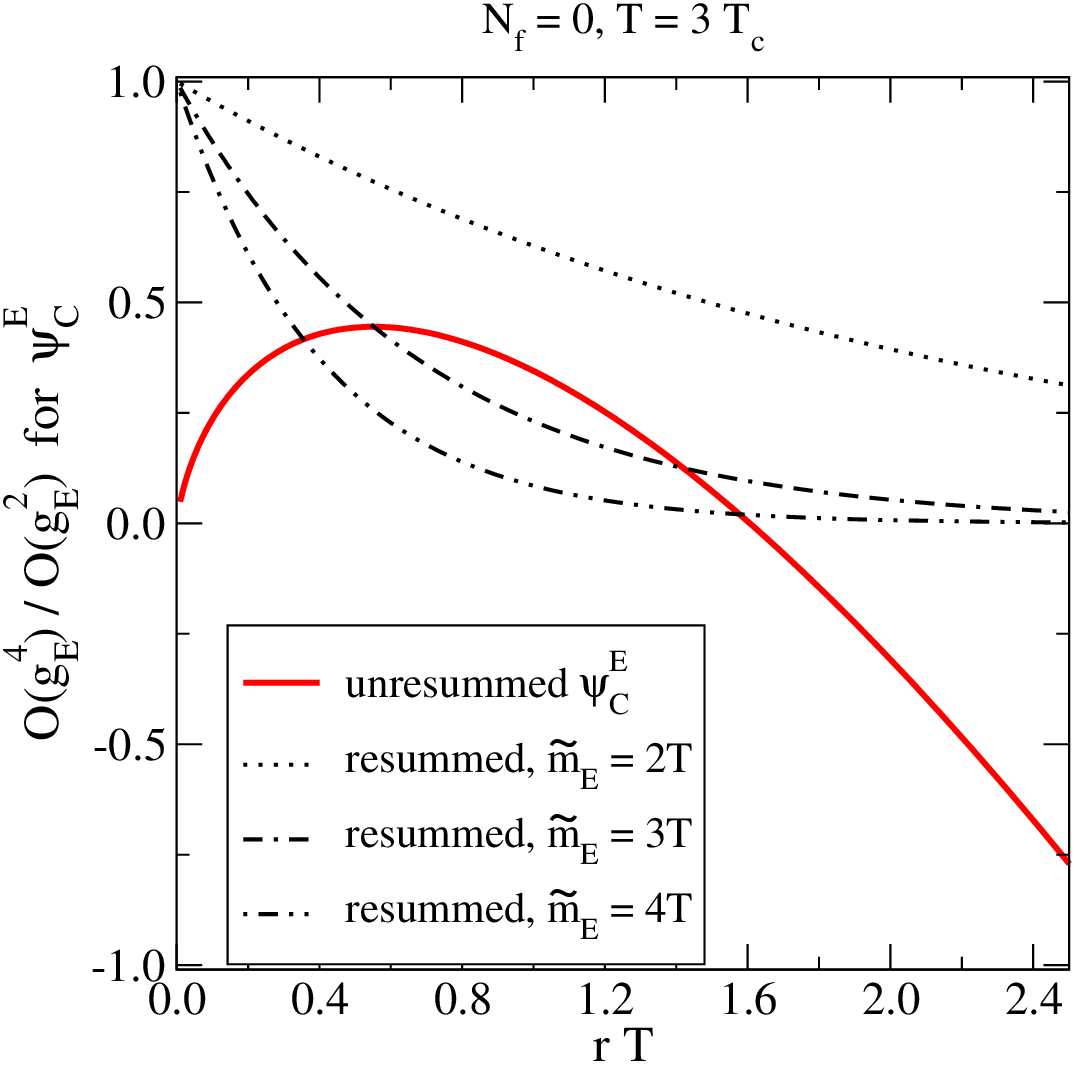}%
}


\caption[a]{\small
The $\rmO(\gE^4)$
correction to $\psi_\rmii{C}^\rmii{E}$ (the expression with 
curly brackets in \eq\nr{psi_E_final}) over the $\rmO(\gE^2)$
term (the first $r$-dependent term in \eq\nr{psi_E_final}), 
labelled as the ``unresummed $\psi_\rmii{C}^\rmii{E}$''; 
together with the same ratio
emerging from the procedure of \eqs\nr{resum_mass}--\nr{tmE}, 
labelled as the ``resummed'' method. At small distances, 
the unresummed method shows better apparent convergence, 
particularly if $\tilde{m}_\rmii{E}$ is small 
as would be preferred by \fig\ref{fig:psiC} 
(where $\tilde{m}_\rmii{E}> 2 T$ would lead to stronger screening 
and a larger deviation from lattice data); 
however, the unresummed method must not be used at  $r \gg 1/T$. 
}

\la{fig:test}
\end{figure}

In \fig\ref{fig:test}, we compare the magnitude of the $\rmO(\gE^4)$
correction with that of the $\rmO(\gE^2)$ term, with the former one
computed either as in \eq\nr{psi_E_final}, 
or as in \eqs\nr{resum_mass}--\nr{tmE}. 
It appears that if 
we focus on small distances, $r\lsim 2/T$, then the ``unresummed''
method, i.e.\ \eq\nr{psi_E_final}, shows better convergence. 
Therefore in the following we mostly concentrate on this case 
although, as the discussions above and in \fig\ref{fig:test} show, 
the unresummed result cannot be extrapolated to large 
distances, $r\gsim 2/T$. 

%
\subsection{Summary and comparison with literature}

For moderate distances, $r \ll \pi/g^2 T$, we can 
write down a ``full result'' by adding up \eqs\nr{psi_C_inter} 
and \nr{psi_E_final}: 
\ba
 & & \hspace*{-0.3cm} 
 \ln\biggl( \frac{\psi_\rmii{C}(r)}{|\psi_\rmii{P}|^2}\biggr)
  \approx 
 \frac{g^2 C_F  \exp(-\mE r)}{4\pi T r}
 \biggl\{1 + \frac{g^2}{(4\pi)^2}
    \biggl[
     \frac{11\Nc}{3} \biggl( 
                              L_\bo + 1 \biggr)  
    - 
     \frac{2\Nf}{3} 
   \biggl(
      L_\fe - 1 
   \biggr)  
   \biggr] \biggr\}
 \nn[3mm] & & \; 
 + \;  
 \frac{g^4 C_F \CA \exp(-\mE r)}{(4\pi)^2}
 \biggl[ 
   2 - \ln(2\mE r) - \gammaE + e^{2 \mE r} E_1 (2 \mE r)
 \biggr]
 - \frac{g^4 C_F \CA}{(4\pi)^2}
   \frac{\exp(-2 \mE r)}{8 T^2 r^2} 
 \nn[3mm] & & \; 
  + \; \frac{g^4 C_F \CA}{(4\pi)^2} 
 \biggl[ 
   \frac{1}{12 T^2 r^2}
  + \frac{{\rm Li}_2(e^{-4\pi T r})}{(2\pi T r)^2}
  + \frac{1}{\pi T r} 
    \int_1^{\infty} \! {\rm d} x \, 
    \biggl(\frac{1}{x^2} - \frac{1}{2 x^4} \biggr)
    \ln\Bigl( 1 - e^{-4\pi T r x} \Bigr)
 \biggr]
 \nn[3mm] & & \; 
 + \; \frac{g^4 C_F \Nf}{(4\pi)^2} 
 \biggl[ 
  \frac{1}{2 \pi T r} 
    \int_1^{\infty} \! {\rm d} x \, 
    \biggl(\frac{1}{x^2} - \frac{1}{x^4} \biggr)
    \ln\frac{ 1 + e^{-2\pi T r x} }{ 1 - e^{-2\pi T r x} } 
 \biggr] + \rmO(g^5) 
 \;, \hspace*{1cm} \la{psi_C_final}
\ea
where $L_\bo, L_\fe$ are as defined in \eq\nr{log_defs} and
$E_1, {\rm Li}_2$ in \eqs\nr{E1}, \nr{Li2}.
This result is renormalization group invariant up to $\rmO(g^5)$.  
For large distances, the square bracket term on the 
second line is to be omitted,  
and the exponential function on the first line is to be 
replaced with $\exp(- \tilde{m}_\rmii{E} r)$, with some 
non-perturbative $\tilde{m}_\rmii{E}$. 
The uncancelled power-law term on the third line implies 
that the singlet free energy  dies away at large distances 
slower than gauge-invariant correlations.

\begin{figure}[t]


\centerline{%
\epsfysize=7.5cm\epsfbox{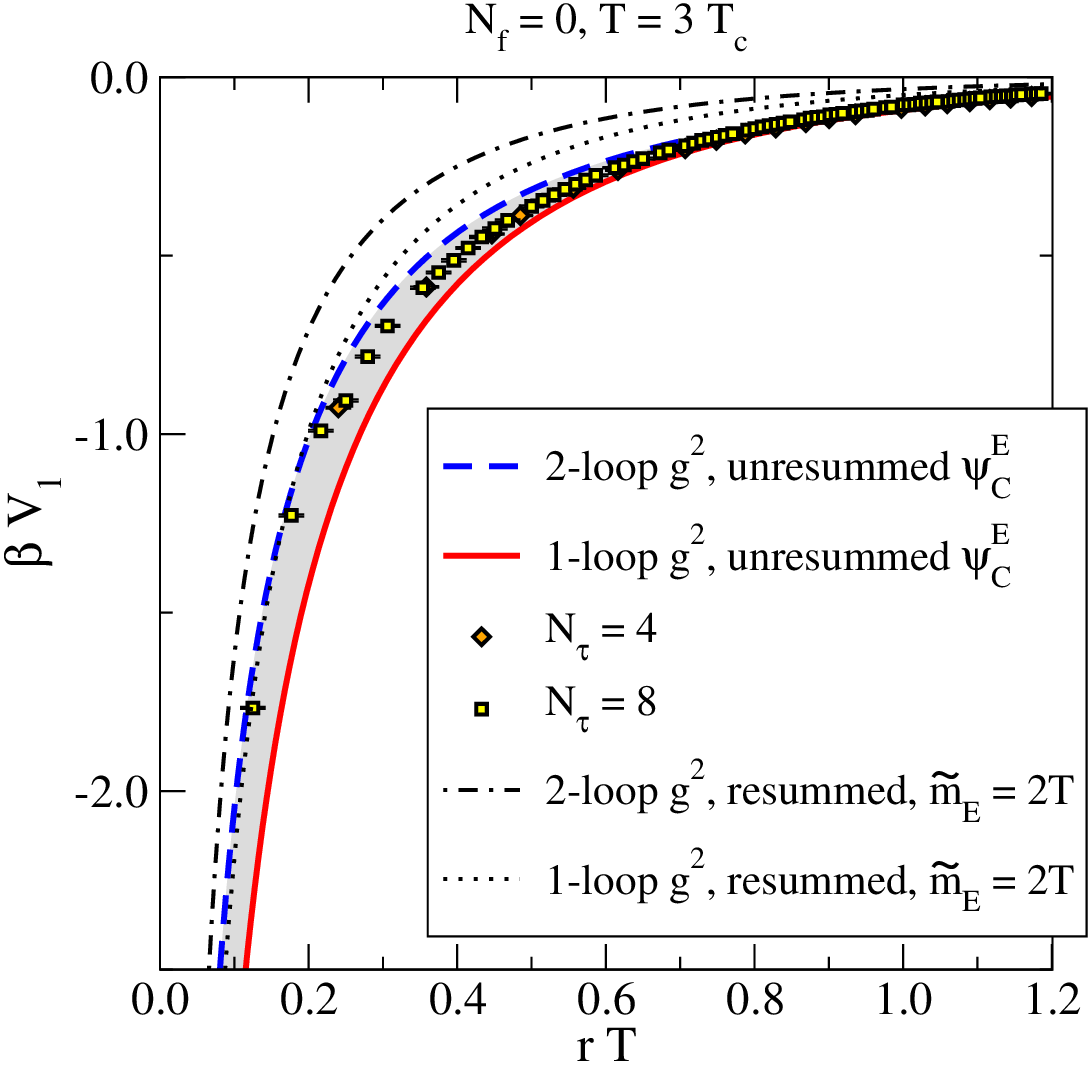}%
~~\epsfysize=7.5cm\epsfbox{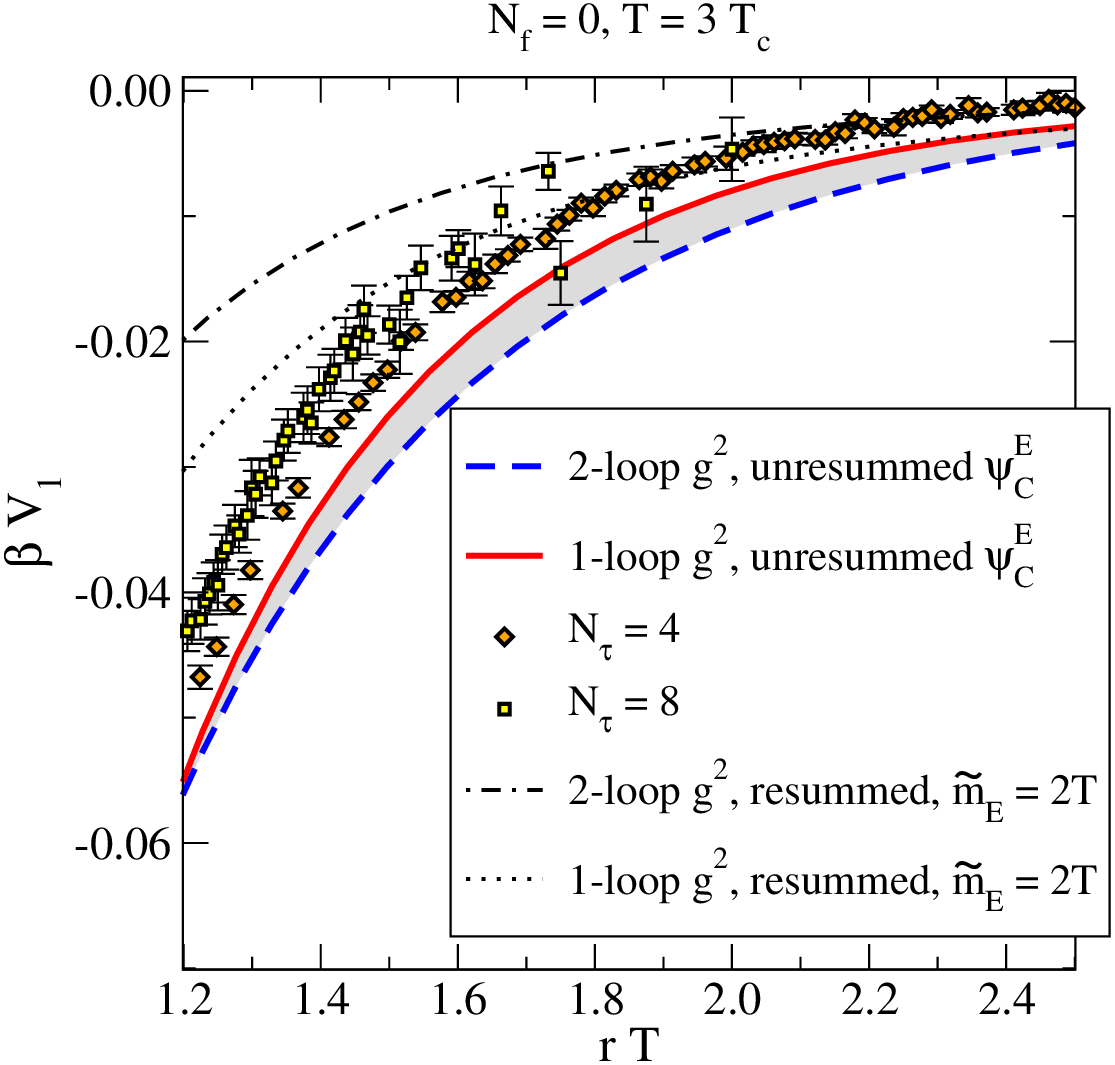}%
}


\caption[a]{\small
The singlet potential in Coulomb gauge from \eq\nr{psi_C_final}
[$
 \beta V_1 \equiv - \ln( {\psi_\rmii{C}} / {|\psi_\rmii{P}|^2})
$], 
for $\Nf = 0$, at $T = 3.75 \Lambdamsbar$ 
($T \approx 3 \Tc$), at small (left) and large (right) distances.  
The band corresponds to variations of the 
gauge coupling and $\mE$ as explained in the
caption of \fig\ref{fig:psiP}. 
As explained in \fig\ref{fig:test}, ``unresummed'' results can only
be applied at short distances, 
``resummed'' ones only at large distances. 
The lattice data, labelled by $N_\tau$, is from ref.~\cite{q_singlet}
(the spatial lattice size was kept fixed at $32^3$). 
}

\la{fig:psiC}
\end{figure}

In \fig\ref{fig:psiC} we compare \eq\nr{psi_C_final} with $\Nf = 0$ 
lattice data from ref.~\cite{q_singlet}. The parameters have been 
fixed as in \eq\nr{eff_coup}, and also more elaborately as explained
in the caption.  We observe good agreement between 
our result and the non-perturbative data, if the unresummed
form of \eq\nr{psi_E_final} is used at short
distances, and the resummed form of \eqs\nr{resum_mass}--\nr{tmE} at large 
distances. (Unfortunately the latter expression involves an unknown
parameter, $\tilde{m}_\rmii{E}$, so the test is less stringent
at large distances.) We have repeated the comparison at $T\approx 12 \Tc$, 
and the agreement remains good, despite the band becoming narrower
(cf.\ \fig\ref{fig:psiC_12Tc}). 
Such a nice agreement for $\psi_\rmii{C}$ 
even at $T\approx 3 \Tc$ is perhaps somewhat 
surprising, given that according to \fig\ref{fig:psiP} 
higher-order perturbative corrections to $\psi_\rmii{P}$ 
could still to be significant 
in this temperature range. (Formally, $\psi_\rmii{P}$ can be obtained
from the $T$- and 
$r$-dependent part of $\psi_\rmii{C}$ by setting $r\to 0$, cf.\ 
\eqs\nr{psi_P_full}, \nr{psi_C_full}, and it can indeed be observed from 
\fig\ref{fig:psiC}(left) that some tendency towards a discrepancy starts
to form in this limit.)

\begin{figure}[t]


\centerline{%
\epsfysize=7.5cm\epsfbox{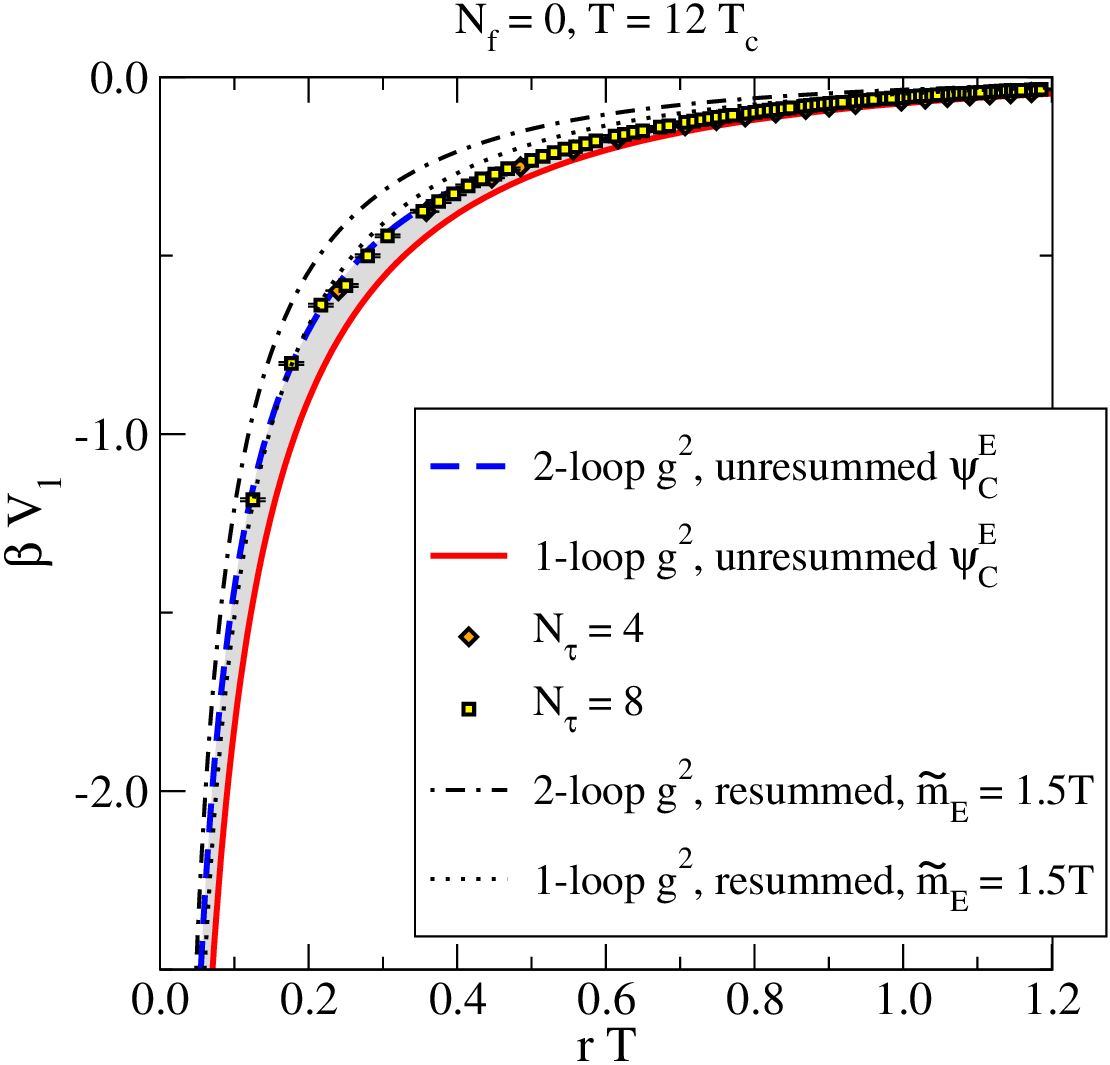}%
~~\epsfysize=7.5cm\epsfbox{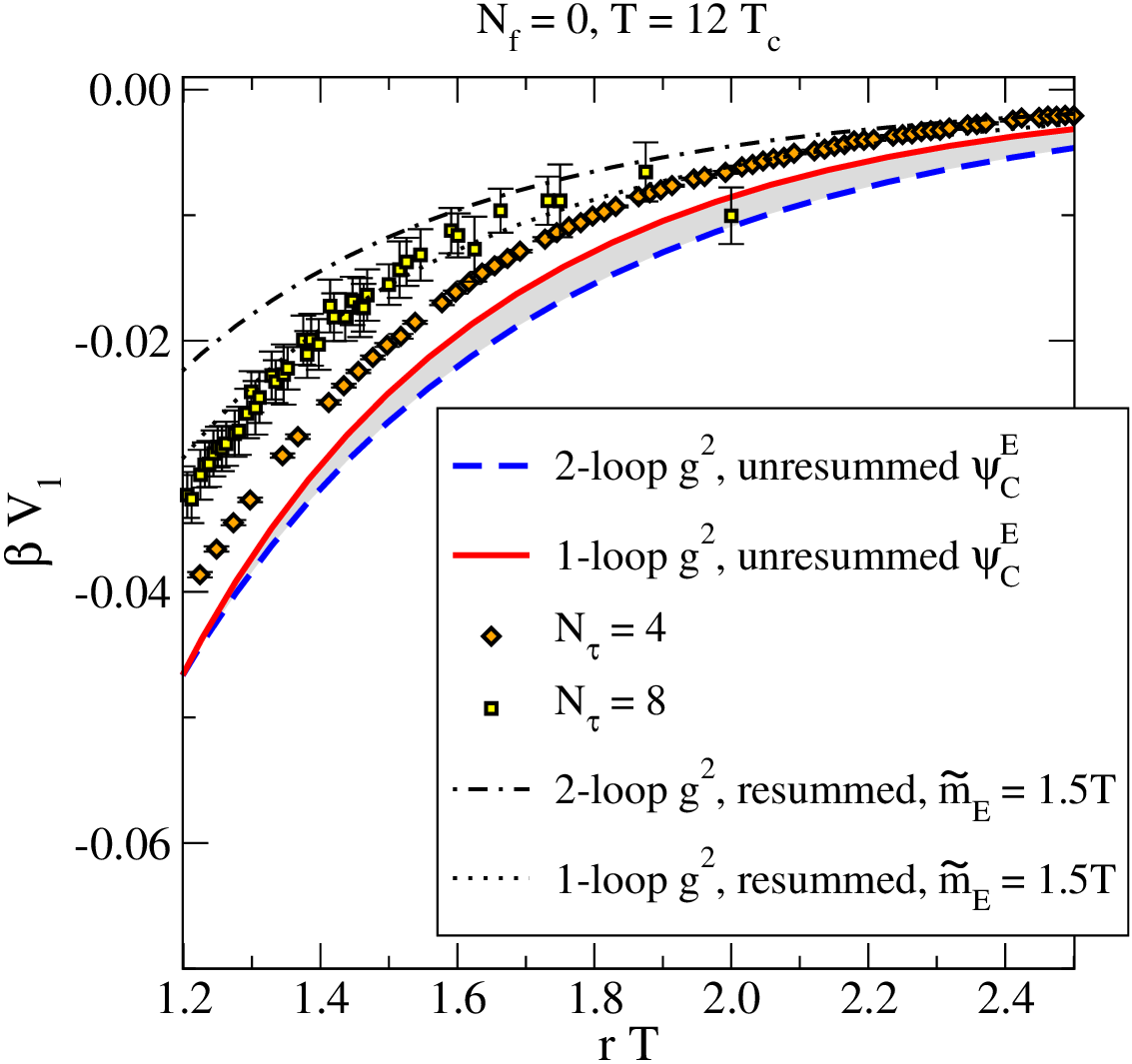}%
}


\caption[a]{\small
Like \fig\ref{fig:psiC} but for $T = 15 \Lambdamsbar$ 
($T \approx 12 \Tc$). The parameter $\tilde{m}_\rmii{E}$ has 
been scaled down roughly by an amount corresponding to the running 
of the gauge coupling~\cite{gE2}.
}

\la{fig:psiC_12Tc}
\end{figure}

It is interesting to compare the present results with those 
in ref.~\cite{hbm}, where the short-distance spatial correlators 
related to gauge-invariant scalar and pseudoscalar densities were 
measured. The authors observed stronger correlations than 
indicated by the {\em leading-order} perturbative predictions
(in the language of the static potential, this would correspond
to less screening). It would
be interesting to see whether pushing the perturbative determination
of those observables
to the same level as here would allow to make the match 
as good as that in \fig\ref{fig:psiC}.
 
%
\section{Singlet free energy in a general covariant gauge}
\la{se:xi}

The singlet free energy in a general covariant gauge 
is defined in analogy with \eq\nr{Vsinglet}, 
\be
 \psi_{\tilde\xi}(r) \equiv 
 \frac{1}{\Nc}
 \langle \tr[P_\vec{r} P^\dagger_\vec{0} ] \rangle_{\tilde\xi}
 \;. \la{Vxi}
\ee
The graphs are the same as in \fig\ref{fig:Coulomb}, and 
so is the general unresummed result of \eq\nr{psi_C_full}.
The difference is that for 
the gauge field propagator we now insert 
\be
 G_{\mu\nu}(Q) = \frac{\delta_{\mu\nu}}{Q^2} + 
 \tilde \xi \frac{Q_\mu Q_\nu}{Q^4}
 \;, \la{G_xi}
\ee
while the self-energy has the form 
\ba
 \Pi_{00}(0,\vec{k}) & = & 
 g^2 \CA 
 \Tint{Q}
 \biggl\{ 
   \frac{D-2}{Q^2} -  \frac{2[k^2 + (D-2) q_n^2]}{Q^2(Q+K)^2}
    \nn & & \; + 
   \frac{\tilde\xi  k^2}{Q^4} - 
    \frac{\tilde\xi k^2 [ k^2 + 2 q_n^2 ]}{Q^4(Q+K)^2} 
   - \frac{\tilde\xi^2 k^4 q_n^2}{2 Q^4(Q+K)^4}
 \biggr\}
   \nn &  + &  
 g^2 \Nf \Tint{\{ Q \} }
 \biggl\{ 
   -\frac{2}{Q^2} + \frac{k^2 + 4 q_n^2}{Q^2(Q+K)^2}
 \biggr\} 
 \;, \la{Pi_xi}
\ea
where $K\equiv (0,\vec{k})$. Expanding in $k^2$
following the philosophy of \eq\nr{strategy}, the self-energy behaves as 
\ba
 & & \hspace*{-1.0cm} 
 \Pi_{00}(0,\vec{k})  =   \mE^2 
 \nn & & \hspace*{-0.5cm}
 +  \frac{g^2 k^2}{(4\pi)^2}
 \biggl\{   
  -\frac{\Nc}{6} \biggl[
   ({10 - 3\tilde\xi})
    \biggl(\frac{1}{\epsilon} + L_\bo \biggr)
    + {6\tilde\xi - 2} 
   \biggr] + 
   \frac{2\Nf}{3} 
   \biggl(
     \frac{1}{\epsilon} + L_\fe - 1 
   \biggr)  
 \biggr\}
 + \rmO\biggl( \frac{k^4}{T^2}\biggr)
 \;, \la{Pi_xi_exp}
\ea
where terms of $\rmO(k^4/T^2)$ are ultraviolet finite. 
The fermionic part here has a structure familiar from \eq\nr{psi_C_inter}.

While in principle we could go on as before, working out 
the contribution of $\rmO(k^4/T^2)$ in detail, it seems to 
us that the result is not particularly interesting. This can 
be seen already at short distances, $r \pi T \ll 1$, focussing
on the divergences, as we will now do.

Listing only the $1/\epsilon$-poles of dimensional regularization, 
the renormalization of the bare gauge coupling (cf.\ \eq\nr{gBare}) from 
the first graph of \fig\ref{fig:Coulomb} yields the contribution
\be
 \delta \psi_{\tilde\xi}^{(1)}  = 
 \frac{g^4 C_F}{(4\pi)^3 T r}
    \biggl[ -\frac{11 \Nc}{3} + \frac{2\Nf}{3}
    \biggr]
    \biggl[
       \frac{1}{\epsilon}
    + \rmO\bigl(1  \bigr) \biggr]
   \;. 
\ee
The second and third graphs produce 
\be
 \delta \psi_{\tilde\xi}^{(2-3)}  = 
 \frac{g^4 C_F}{(4\pi)^3 T r}
    \biggl[ \bigl(4 - 2 \tilde\xi \bigr)\Nc
    \biggr]\biggl[\frac{1}{\epsilon}
    + \rmO\bigl(1  \bigr) \biggr] 
 \;, 
\ee
while the fourth and fifth graphs are ultraviolet finite. 
The sixth graph yields
\be
 \delta \psi_{\tilde\xi}^{(6)}  = 
 \frac{g^4 C_F}{(4\pi)^3 T r}
    \biggl[ \fr32 \tilde\xi \Nc
    \biggr]\biggl[\frac{1}{\epsilon}
    + \rmO\bigl(1  \bigr) \biggr]
 \;, 
\ee
and the seventh graph, inserting the expansion 
in \eq\nr{Pi_xi_exp}, amounts to 
\be
 \delta \psi_{\tilde\xi}^{(7)}  = 
 \frac{g^4 C_F}{(4\pi)^3 T r}
    \biggl[ \biggl(\fr53 - \frac{\tilde\xi}{2} \biggr)\Nc
    -\fr23 \Nf 
    \biggr]\biggl[\frac{1}{\epsilon}
    + \rmO\bigl(1  \bigr) \biggr]
 \;.
\ee
Adding up, we observe that divergences do not cancel, 
unlike in Coulomb gauge, but that the results sum up to 
\be
 \psi_{\tilde\xi} = 1 + \frac{g^2 C_F}{4\pi T r}
 + \frac{g^4 C_F \CA}{(4\pi)^3 T r}
 (2-\tilde\xi)\biggl[ \frac{1}{\epsilon}
    + \rmO\bigl(1  \bigr) \biggr]
 \;. \la{div_sum}
\ee 
In principle we could ``renormalize'' the result by writing 
\ba
 \ln\biggl( 
 \frac{\psi_{\tilde\xi}(r)}{|\psi_\rmii{P}|^2} \biggr) 
 & \approx &  \mathcal{Z}_{\tilde\xi} \times 
 \biggl\{ \frac{g^2 C_F}{4\pi T r} 
  + \frac{g^4 C_F \CA}{(4\pi)^3 T r}
 (2-\tilde\xi) 
    \rmO\bigl(1  \bigr) \biggr\}
 \;, \quad \la{pre_Z_xi} \\ 
 \mathcal{Z}_{\tilde\xi} & = &  1 + 
 (2 - \tilde\xi) \frac{g^2 \CA}{(4\pi)^2\epsilon} + \rmO(g^4)
 \;, \la{Z_xi}
\ea
however the form is not that of known endpoint divergences
(see, e.g., ref.~\cite{Aoyama:1981ev}) which are additive, 
rather than ``correcting'' the renormalized $g^2$; in addition, the 
finite parts of the potential still remain 
gauge dependent, even at short distances $r\pi T \ll 1$, 
because of the $\rmO(1)$ term in \eq\nr{pre_Z_xi}. 
Therefore the result does, in general, not match the zero-temperature 
potential. The situation is analogous to that in the next section,
where it will be discussed somewhat more explicitly
(cf.\ \eqs\nr{Z_W}, \nr{psi_W_zeroT}). 

Let us point out that, compared with the conventional
zero-temperature static potential
where divergences do cancel~\cite{wf}, the only difference is with 
the second and third topologies 
in \fig\ref{fig:Coulomb}. At zero temperature, 
the time direction is infinite and the Wilson loop has a finite
extent within it; at finite temperature, the time direction is 
finite and the Polyakov lines wrap all the way around the time direction. 
This turns out to lead to a difference for these particular topologies. 

%
\section{Cyclic Wilson loop}
\la{se:cyclic}

%
\subsection{Basic setup}

With the objects in \eqs\nr{Pdef} and \nr{Wdef},
the cyclic Wilson loop is defined as 
\be
 \psi_\rmii{W}(r) \equiv
 \frac{1}{\Nc}
 \langle \tr[P_\vec{r} W_0 P^\dagger_\vec{0} W^\dagger_0] \rangle  
 \;, \la{psi_W_def}
\ee
where we made use of periodic boundary 
conditions (i.e.\ $W_\beta = W_0$). In an Abelian theory 
$\psi_\rmii{W}$ 
would agree with $\psi_\rmii{C}$ of \eq\nr{Vsinglet}, whereas
in the non-Abelian case $\psi_\rmii{W}$ can be viewed
as a gauge-invariant ``completion'' of $\psi_\rmii{C}$.

%
\begin{figure}[t]
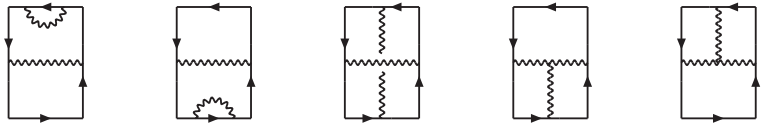


\begin{eqnarray*}
&& 
 \hspace*{-2cm}
 \CyclA \quad 
 \CyclB \quad 
 \CyclC \quad 
 \CyclD \quad 
 \CyclE
\end{eqnarray*}

\caption[a]{\small 
The additional graphs relevant for the cyclic Wilson loop, 
supplementing those in \fig\ref{fig:Coulomb}, in the class
of gauges where the gluon propagator has no components mixing
time and space indices, i.e.\ $G_{0i}(P) = 0$. Time runs 
vertically and space horizontally.} 
\la{fig:cyclic}
\end{figure}
%

In general, the insertion of the spacelike Wilson lines into 
the observable leads to a huge proliferation of graphs compared with 
those in \fig\ref{fig:Coulomb}. On the other hand, the fact that both
$W_0$ and $W_0^\dagger$ appear, i.e.\ that the (untraced) Polyakov loops
are really connected by a Wilson line in the adjoint 
representation, also means that there are many cancellations. 
Even so, quite a number of diagrams remains. To simplify the task somewhat, 
we have restricted to special gauges in this case, 
namely to those where the gluon propagator has no components mixing
time and space indices, i.e.\ $G_{0i}(Q) = 0$; this class includes, 
in particular, the Coulomb and the Feynman gauges. With these provisions, 
only the graphs in \fig\ref{fig:cyclic} 
give an additional contribution. Choosing for 
convenience the coordinates so that $\vec{r}$ points in the $z$-direction, 
the general (unresummed) result can be written as 
\ba 
 \psi_\rmii{W}(r) & = & \psi_\rmii{C}(r) + g^4 C_F \CA 
 \biggl[
  \int_{\vec{k}} e^{i \vec{k}\cdot\vec{r}} G_{00}(0,\vec{k})  
   \int_{\vec{q}}
   \Bigl(e^{i \vec{q}\cdot \vec{r}} - 1 \Bigr)
   \sum_{q_n} \frac{G_{zz}(q_n,\vec{q})}{q_z^2}
  \nn & & \;\hspace*{0.5cm} + \; 
   \int_{\vec{k},\vec{q}} \Bigl( e^{-i \vec{q}\cdot \vec{r}} - 
   e^{i \vec{k}\cdot \vec{r}}\Bigr)
      G_{00}(0,\vec{k}) G_{00}(0,\vec{q}) 
      G_{zi}(0,\vec{q+k}) \frac{k_i - q_i}{k_z+q_z} 
 \biggr]
 \;, \hspace*{0.5cm} \la{psi_W_full}
\ea
where $\psi_\rmii{C}$ refers to the general result in \eq\nr{psi_C_full}.
It can be verified, by taking the Coulomb and Feynman gauge results 
for $\psi_\rmii{C}$ from \ses\ref{se:Coulomb}, \ref{se:xi}, respectively, 
as well as the corresponding propagators from \eqs\nr{G_C}, \nr{G_xi}, 
that the expression in \eq\nr{psi_W_full} 
is indeed gauge independent. (In fact, we will
demonstrate the gauge independence of the EQCD part 
of the expression around \eq\nr{psi_W_E_expl} below.) 
It is also trivial to see that formally 
the new contribution in \eq\nr{psi_W_full} vanishes for $r=0$, 
like the structures 
of \eq\nr{psi_C_full}.

Inserting the Coulomb gauge propagator from \eq\nr{G_C} and carrying out
some changes of integration variables, we can write the unresummed result as
\ba
 \psi_\rmii{W}(r) & = & \psi_\rmii{C}(r) + g^4 C_F \CA 
 \int_{\vec{k,q}} \frac{e^{i \vec{k}\cdot\vec{r}}}{k^2} 
 \biggl\{
  \sum_{q_n} 
  \biggl[ 
    \frac{e^{i q_z r} -1}{q_z^2 Q^2} 
    + \frac{1}{q^2 Q^2}\biggl( 1 - \frac{k^2}{(k+q)^2} \biggr)
  \biggr] 
 \nn & & \; \hspace*{2cm} 
 - \frac{2}{q^2(k+q)^2}
  \biggl[
    \frac{k_z-q_z}{k_z+q_z} + \frac{q^2 - k^2}{(k+q)^2} 
  \biggr]
 \biggr\}
 \;. \la{psi_W_expl}
\ea
Now it is understood that
the Coulomb gauge $\psi_\rmii{C}$ 
from \eq\nr{psi_C_expl} is to be inserted.  

%
\subsection{Short-distance limit} 

Let us start by inspecting the new terms in \eq\nr{psi_W_expl}
at short distances, in analogy with \se\ref{psi_C_short}.
The latter row in \eq\nr{psi_W_expl} only contains the Matsubara
zero mode and, compared with the contribution from the sum,  
leads for dimensional reasons to a contribution suppressed 
by $r T$ at small distances (once $\beta$ is factored out). 
The zero-temperature integrals corresponding to the structures
on the first row can be carried out, 
\ba
 \int_Q \frac{e^{i q_z r} -1}{q_z^2 Q^2} 
 & = & 
 \frac{1}{8\pi^2}
 \biggl\{ 
   \frac{1}{\epsilon} + 
   2 \biggl[
     \ln\biggl( \frac{\bmu r}{2} \biggr) + \gammaE + 1  
   \biggr]
 \biggr\} 
 \;, \la{sd_W_1} \\ 
 \int_Q \frac{1}{q^2 Q^2}\biggl( 1 - \frac{k^2}{(k+q)^2} \biggr)
 & = & 
 \frac{1}{8\pi^2}
  \biggl(
   \frac{1}{\epsilon} + 
     \ln\frac{\bmu^2}{4 k^2} 
  \biggr)
 \;. \la{sd_W_2}
\ea
The divergences add up; given that $\psi_\rmii{C}$ 
was finite after charge renormalization, 
there is nothing more available that could cancel them. We may, 
nevertheless, represent their effect by introducing a ``renormalization 
factor'' $\mathcal{Z}_\rmii{W}$ in analogy with \eqs\nr{pre_Z_xi}, \nr{Z_xi}, 
\ba
 \ln\biggl( 
 \frac{\psi_\rmii{W}(r)}{|\psi_\rmii{P}|^2} \biggr) 
   & = &  \mathcal{Z}_\rmii{W}\, g^2 C_F \beta \int_\vec{k} \! 
 \frac{e^{i\vec{k}\cdot\vec{r}}}{k^2} + \rmO(g^4)
 \;, \la{Z_W_def} \\ 
 \mathcal{Z}_\rmii{W} 
   & = &  1 +  \frac{4 g^2 \CA}{(4\pi)^2\epsilon} + \rmO(g^4)
 \;, \la{Z_W}
\ea
but this does not have the form normally related to 
the cusps that appear in the observable~\cite{Polyakov:1980ca}
(the corresponding divergences are additive, 
rather than ``correcting'' the renormalized $g^2$). 
In addition, there is an ambiguity concerning whether
the Fourier transform in \eq\nr{Z_W_def} is to be understood 
in $3-2\epsilon$ or in 3 dimensions. Preferring therefore
to write down a bare expression, we can summarize the 
short-distance behaviour as 
\be
 \psi_\rmii{W} (r)  
 \; 
 \stackrel{r \pi T \ll 1}{\approx}
 \;
 \psi_\rmii{C}(r)
 + \frac{g^4 C_F \CA\beta}{(4\pi)^2 }
 \int_\vec{k} \! 
 \frac{e^{i\vec{k}\cdot\vec{r}}}{k^2}
 \biggl[ 4\biggl( \frac{1}{\epsilon} + 
    \ln\frac{\bmu^2}{4 k^2}   + 1 \biggr)  
 \biggr]
 \;,  \la{psi_W_zeroT}
\ee
where we made use of footnote~\ref{f4} in order to 
rephrase all the logarithmic dependence in $\vec{k}$-space. 
The expression in \eq\nr{psi_W_zeroT} is gauge-independent; however, 
it {\em does not} go over into 
the zero-temperature potential at short distances.

%
\subsection{Hard-mode contribution}

In order to determine the behaviour of $\psi_\rmii{W}$
at larger distances, $r\sim 1/\pi T$,
we treat the new terms in \eq\nr{psi_W_expl} according to 
the strategy around \eq\nr{strategy}. Two of the sum-integrals
are by now familiar, and given by \eqs\nr{si6}, \nr{rint_3}; the new one
is given by \eq\nr{rint_7}, and we get
\ba
 & & \hspace*{-1cm} 
 \Bigl[(\psi_\rmii{W}(r))_\rmi{QCD}\Bigr]_\rmi{unresummed} \; = \; 
 \Bigl[(\psi_\rmii{C}(r))_\rmi{QCD}\Bigr]_\rmi{unresummed} 
 \nn[3mm] & & + \; 
 \frac{g^4 C_F \CA \beta}{(4\pi)^2} 
 \int_{\vec{k}} \frac{e^{i \vec{k}\cdot\vec{r}}}{k^2}  
     \biggl\{
     4 \biggl( \frac{1}{\epsilon} + L_\bo + 1 \biggr)  
   \biggr\}
  \nn[3mm] & & \; 
  + \frac{g^4 C_F \CA}{(4\pi)^2} 
 \biggl[ 
  - \frac{1}{12 T^2 r^2}
 + \frac{2 {\rm Li}_2(e^{-2\pi T r})}{(2\pi T r)^2}
 + \frac{1}{\pi T r} \int_1^\infty \! \frac{{\rm d} x}{x^2} 
 \ln\Bigl( 1 - e^{-2\pi T r x} \Bigr) 
 \biggr]
  \nn[3mm] & & + \; 
    g^4 C_F \CA \times \mbox{(zero mode contribution)}
 \;. \la{psi_W_inter}
\ea
Here we already replaced 
the sum over the exponential integral through
a simple integral representation, cf.\ \eq\nr{int_rep}.

At short distances, $\pi T r \ll 1$, 
the complicated square bracket expression in \eq\nr{psi_W_inter}
goes over into
\be
 \frac{g^4 C_F \CA}{(4\pi)^3 T r}
 \biggl[
    8 \ln (2\pi T r) 
 \biggr] 
 \;.
\ee
Combining with the logarithmic term from 
the first row (with $L_\bo$ inserted from \eq\nr{log_defs}), 
we reproduce the result of \eq\nr{psi_W_zeroT}. 
At large distances, on the other hand, 
the power-law term in \eq\nr{psi_W_inter}
cancels against that in \eq\nr{problem}. 
Therefore, apart from a disconnected part like
in \eq\nr{asympt}, $\psi_\rmii{W}(r)$ is exponentially suppressed
at large distances.

%
\subsection{Soft-mode contribution}

We denote the EQCD contribution 
to $\psi_\rmii{W}$ by $\psi_\rmii{W}^\rmii{E}$. 
Within EQCD, the new graphs in \fig\ref{fig:cyclic} amount to 
an evaluation of the expectation value (see also ref.~\cite{sn2})
\be
 \psi_\rmii{W}^\rmii{E}(r) = 
 \psi_\rmii{C}^\rmii{E}(r) + \frac{g^2 \beta^2 \mathcal{Z}_1^2}{\Nc}
 \Bigl\langle
   \mathcal{Y}_0^2\tr [ A_0(\vec{r}) W_0 A_0(\vec{0}) W_0^\dagger ]
 - \tr [ A_0(\vec{r}) A_0(\vec{0}) ]
 \Bigr\rangle + \rmO(g^5)
 \;, \la{psi_W_E}
\ee
where the subtraction corresponds to the part already 
included in $\psi_\rmii{C}^\rmii{E}$. We have introduced 
another ``$\mathcal{Z}$-factor'', this time denoted by $\mathcal{Y}_0$, 
related to the fact that the spatial Wilson lines within EQCD might
differ in normalization from those in QCD. 
The new contributions we
are interested in are $\rmO(g^4)$, so that $\mathcal{Z}_1$
can be set to unity and $\gE^2$ can be set to $g^2$
in the evaluation of \eq\nr{psi_W_E}. 
A straightforward computation leads to 
\ba
 \psi_\rmii{W}^\rmii{E}(r) \!\! & = & \!\! 
 \psi_\rmii{C}^\rmii{E}(r) 
 +  g^2 C_F \beta \,
 \Bigl(\mathcal{Y}_0^2-1\Bigr)\! 
 \int_{\vec{k}} \frac{e^{i\vec{k}\cdot\vec{r}}}{k^2 + \mE^2}
 \nn &+  & \; 
  g^4 C_F \CA 
 \int_{\vec{k}} e^{i \vec{k}\cdot\vec{r}}
 \int_{\vec{q}} 
 \biggl[ 
  \frac{1}{k^2 + \mE^2} \frac{e^{i q_z r}-1}{q_z^2 q^2} 
  + \nn &  & \;  +  \; 
   \frac{2(q_z - k_z)}{q_z + k_z}
     \frac{1}{(q+k)^2 (k^2 + \mE^2)(q^2 + \mE^2)}
  + \frac{\tilde\xi}{q^4}
  \biggl( 
     \frac{1}{k^2 + \mE^2} - \frac{1}{(k+q)^2 + \mE^2}
  \biggr) 
 \biggr] 
 \;,  \nn \la{psi_W_E_expl}
\ea
where we have for completeness kept a general gauge parameter.
It is easy to check now that the $\tilde\xi$-dependent 
part cancels against the self-energy contribution to 
$\psi_\rmii{C}^\rmii{E}$
from the 1st row of \eq\nr{resum_mass}, 
with the self-energy inserted from \eq\nr{Pi_E}, 
so the result is indeed gauge independent. 

On the other hand, if we take the Coulomb gauge result as a starting
point, then all terms in \eq\nr{psi_W_E_expl} need to be kept, with 
the value $\tilde\xi = -1$. The term on the 2nd row of \eq\nr{psi_W_E_expl}
is factorized and contains the integral
\be
 \int_{\vec{q}} \frac{e^{i q_z r} - 1}{q_z^2 \, q^2} = 
 - \frac{r}{8\pi}
 \biggl\{
   \frac{1}{\epsilon_\rmii{UV}} + 
   2 \Bigl[ 
     \ln(\bmu r) + \gammaE - 1
   \Bigr] 
 \biggr\}
 \;, \la{i1} 
\ee
multiplied by the usual 
$
 g^4 C_F \CA  \int_{\vec{k}} e^{i\vec{k}\cdot\vec{r}}/(k^2 + \mE^2) = 
 g^4 C_F \CA \exp(-\mE r)/ 4\pi r + \rmO(\epsilon)
$ (cf.\ \eq\nr{amb1}).
It is important to stress that, as underlined by the notation, 
the divergence in \eq\nr{i1} has an ultraviolet origin. This can be seen, 
for instance, by regulating the $q_z$-integral by taking a principal
value and the $q_\bot$-integral by introducing a mass-like regulator:
\ba
  \int_{\vec{q}} \frac{e^{i q_z r} - 1}{q_z^2 \, q^2} 
  & \rightarrow & 
 \int_{-\infty}^{\infty} \! \frac{{\rm d} q_z}{2\pi }
 \, \mathbb{P} \biggl( \frac{e^{i q_z r} - 1}{q_z^2} \biggr)
 \int \! \frac{{\rm d}^{2-2\epsilon} q_\bot}{(2\pi)^{2-2\epsilon}}
 \frac{1}{q_\bot^2 + q_z^2 + \lambda^2} \nn 
 & = & 
 -\int_0^r \! {\rm d}x \, 
 \int_{-\infty}^{\infty} \! \frac{{\rm d} q_z}{2\pi }
 \, \mathbb{P} \biggl( \frac{e^{i q_z x}}{i q_z} \biggr)
 \, \frac{\mu^{-2\epsilon}}{4\pi}
 \biggl[
    \frac{1}{\epsilon_\rmii{UV}}
  + \ln\frac{\bmu^2}{\lambda^2 + q_z^2} + \rmO(\epsilon)
 \biggr]  
 \;. 
\ea
The $q_z$-independent part of the square brackets 
is multiplied by 
\be
  \int_{-\infty}^{\infty} \! \frac{{\rm d} q_z}{2\pi }
 \, \mathbb{P} \biggl( \frac{e^{i q_z x}}{i q_z} \biggr)
 = \frac{1}{2\pi} \int _{-\infty}^{\infty} \! {\rm d} z \, 
 \frac{\sin z}{z} = \fr12
 \;, 
\ee
whereby we recover the divergent part of \eq\nr{i1}.
This divergence is essentially the self-energy correction 
related to an adjoint Wilson line~\cite{ay}.

As far as the other terms in \eq\nr{psi_W_E_expl} go, 
the last one simply yields
\be
 g^4 C_F \CA 
 \int_{\vec{k,q}} e^{i \vec{k}\cdot\vec{r}}
 \frac{1}{q^4}
  \biggl( 
       \frac{1}{(k+q)^2 + \mE^2} - \frac{1}{k^2 + \mE^2}
  \biggr)
 = 
 - \frac{g^4 C_F \CA e^{-\mE r}}{32\pi^2}
 \;. \la{I_simple}
\ee
The remaining (middle) term requires a bit more work; 
some intermediate steps are given in \eqs\nr{i2}--\nr{I_final_2}.
The final result can be written as 
\ba
 \mathcal{I} &=&
    2  \int_{\vec{k}} e^{i \vec{k}\cdot\vec{r}}
   \frac{q_z - k_z}{q_z + k_z}
     \frac{1}{(q+k)^2 (k^2 + \mE^2)(q^2 + \mE^2)}
 \la{I_def} \\ 
 &=& \frac{e^{-\mE r}}{16\pi^2 \mE r}
  \Bigl\{ 
          (1+\mE r) \Bigl[ \ln(2\mE r) + \gammaE \Bigr]
        + (1-\mE r) e^{2 \mE r} E_1(2\mE r)
  \Bigr\}
 \;. \la{I_final}
\ea
It may be noted that the large-distance behaviour of \eq\nr{I_final} 
largely cancels against that in \eq\nr{psi_E_final}.

To summarize, the additional contribution to $\psi_\rmii{W}^\rmii{E}$
is composed of the first new term in \eq\nr{psi_W_E_expl}, together with 
the results from \eqs\nr{i1}, \nr{I_simple}, and \nr{I_final}. 
At small distances the ``old'' part $\psi_\rmii{C}^\rmii{E}$ could be 
taken from its direct evaluation, \eq\nr{psi_E_final}, while at large 
distances the resummed form of \eqs\nr{resum_mass}--\nr{tmE} is preferable.

It remains to fix $\mathcal{Y}_0^2$ in the first new term 
of \eq\nr{psi_W_E_expl}. In order to match the behaviour on the 2nd 
row of \eq\nr{psi_W_inter} in the subtraction--addition step, 
we need to choose
\be
 \mathcal{Y}_0^2 = 1 + \frac{g^2}{(4\pi)^2}
 \biggl[ 4\Nc 
   \biggl( 
     \frac{1}{\epsilon} + L_\bo + 1
   \biggr) 
 \biggr]
 \;.
\ee
It can be seen, however, that there is a potential ambiguity from terms 
of the type $\rmO(1/\epsilon)\times\rmO(\epsilon)$ that this multiplies. 
Here the $\rmO(\epsilon)$-terms come from the massive or 
massless leading-order potential, 
\ba 
 \int \! \frac{{\rm d}^{3-2\epsilon} \vec{k}}{(2\pi)^{3-2\epsilon}}
 \frac{e^{i\vec{k}\cdot\vec{r}}}{k^2 + \mE^2} & = & 
 \frac{e^{-\mE r}\mu^{-2\epsilon}}{4\pi r}
 \biggl\{
   1 + \epsilon 
   \biggl[ 
     \ln \frac{\bmu^2 r}{2 \mE} + \gammaE - e^{2\mE r} E_1(2 \mE r)
   \biggr] 
 \biggr\}
  \;, \la{amb1} \\
 \int \! \frac{{\rm d}^{3-2\epsilon} \vec{k}}{(2\pi)^{3-2\epsilon}}
 \frac{e^{i\vec{k}\cdot\vec{r}}}{k^2 } & = & 
 \frac{\mu^{-2\epsilon}}{4\pi r}
 \biggl\{
   1 + 2 \epsilon 
   \biggl[ 
     \ln (\bmu r) + \gammaE
   \biggr] 
 \biggr\}
  \;, \la{amb2}
\ea  
where we inserted 
$1 = \mu^{-2\epsilon}[1+\epsilon(\ln\frac{\bmu^2}{4\pi} + \gammaE)]$
in order to fix the dimensions. 
It is not clear to us whether 
the $\rmO(1/\epsilon)\times\rmO(\epsilon)$ terms from here 
can have physical significance. 

In any case, after fixing $\mathcal{Y}_0^2$, the terms can be added up. 
The complete result is not particularly transparent, 
and may be ambiguous as just discussed, so we do not write it 
down explicitly; it suffices to say that the sum can be expressed as 
\be
 \ln\biggl( \frac{\psi_\rmii{W}^\rmii{E} (r)}{|\psi_\rmii{P}|^2} 
 \biggr)_\rmi{resummed} =  
 \mathcal{G}_\rmii{DR}\Bigl( \frac{1}{\epsilon},\frac{\bmu}{T}, rT\Bigr)
 \frac{C_F 
 \exp(-\mE r)}{4\pi T r}
 \; - \; \frac{g^4 C_F \CA}{(4\pi)^2}
   \frac{\exp(-2 \mE r)}{8 T^2 r^2} 
 \;. \la{psi_W_E_final}
\ee
The function $\mathcal{G}_\rmii{DR}$, in which the complications are hidden, 
does have a simple expression in certain limits, however, and these 
will be discussed in the next section.

%
\subsection{Summary and comparison with literature}

Adding up the relevant parts of 
\eqs\nr{psi_C_final}, \nr{psi_W_inter} and \nr{psi_W_E_final}, 
we finally get
\ba
 & & \hspace*{-1cm} 
 \ln\biggl( \frac{\psi_\rmii{W} (r)}{|\psi_\rmii{P}|^2} \biggr) \; \approx \; 
 \mathcal{G}_\rmii{DR}\Bigl( \frac{1}{\epsilon},\frac{\bmu}{T}, rT\Bigr) \, 
 \frac{C_F 
 \exp(-\mE r)}{4\pi T r}
 \; - \; \frac{g^4 C_F \CA}{(4\pi)^2}
   \frac{\exp(-2 \mE r)}{8 T^2 r^2} 
 \nn & &  
 \; + \; \frac{g^4 C_F \CA}{(4\pi)^2} 
 \biggl\{ 
 \frac{2{\rm Li}_2(e^{-2\pi T r}) + {\rm Li}_2(e^{-4\pi T r})}{(2\pi T r)^2}
 \nn & & \; \hspace*{1cm}
  +\; \frac{1}{\pi T r} 
    \int_1^{\infty} \! {\rm d} x \, \biggl[ 
    \frac{1}{x^2}\ln\Bigl( 1 - e^{-2\pi T r x} \Bigr) + 
    \biggl(\frac{1}{x^2} - \frac{1}{2 x^4} \biggr)
    \ln\Bigl( 1 - e^{-4\pi T r x} \Bigr)
    \biggr]
 \biggr\}
 \nn & & 
 \; + \; \frac{g^4 C_F \Nf}{(4\pi)^2} 
 \biggl[ 
  \frac{1}{2 \pi T r} 
    \int_1^{\infty} \! {\rm d} x \, 
    \biggl(\frac{1}{x^2} - \frac{1}{x^4} \biggr)
    \ln\frac{ 1 + e^{-2\pi T r x} }{ 1 - e^{-2\pi T r x} } 
 \biggr] 
 + \rmO(g^5) \;. \hspace*{1cm} \la{psi_W_final}
\ea
At small distances, $\mE r \ll 1$, the dominant term of the coefficient 
function $\mathcal{G}_\rmii{DR}$, defined in \eq\nr{psi_W_E_final}, is 
\be
  \mathcal{G}_\rmii{DR}\Bigl( \frac{1}{\epsilon},\frac{\bmu}{T}, rT\Bigr)
  \; \stackrel{\mE r \ll 1}\approx \; g^2 \, 
  \biggl\{ 
     1 + \frac{g^2}{(4\pi)^2} 
   \biggl[ 
     4\Nc \biggl( \frac{1}{\epsilon} + \ln\frac{\bmu^2}{T^2} + \rmO(1) \biggr)
   \biggr]
  \biggr\}
  \;. \la{Gdr_small}
\ee
Though we have not carried out the computation, we could  
expect that in lattice regularization the corresponding
structure goes over into 
\be
  \mathcal{G}_\rmii{lat}\Bigl( \frac{1}{aT}, rT\Bigr)
  \approx g^2 \, 
  \biggl\{ 
     1 + \frac{g^2}{(4\pi)^2} 
   \biggl[ 
     4\Nc \biggl( \frac{1}{a^2T^2} + \rmO(1) \biggr)
   \biggr]
  \biggr\}
  \;, \la{Glat_small} 
\ee
where $g^2$ is a suitably defined {\em renormalized} coupling. 
Therefore, it would appear that 
$
 \mathcal{G}_\rmii{lat} 
$
diverges in the continuum limit. On the other hand, 
at large distances, $\mE r \gg 1$, the dominant term of 
the coefficient function is 
\be
  \mathcal{G}_\rmii{DR}\Bigl( \frac{1}{\epsilon},\frac{\bmu}{T}, rT\Bigr)
  \; \stackrel{\mE r \gg 1}\approx \; g^2 \, 
  \biggl\{ 
     1 - \frac{g^2\Nc T }{8\pi} 
   \biggl[ 
     r \biggl( \frac{1}{\epsilon_\rmii{UV}} + \ln\frac{\bmu^2}{\mE^2} + \rmO(1) \biggr)
   \biggr]
  \biggr\}
  \;, \la{Gdr_large}
\ee
where possible logarithms of $\mE r$ are also included in $\rmO(1)$
(whether such logarithms appear is related to the ambiguities mentioned 
after \eqs\nr{amb1}, \nr{amb2}).\footnote{%
  No logarithms of $r$ appear if the soft contributions to $\psi_\rmii{C}$
  are treated as in \eq\nr{resum_mass} and no terms of the type 
  $\rmO(1/\epsilon)\times\rmO(\epsilon)$ are included, i.e., 
  if we just sum together \eq\nr{i1} and the large-$r$ limit 
  of \eq\nr{I_final}. 
  } 
Apart from the said 
logarithms, \eq\nr{Gdr_large} can be accounted for by a mass correction, 
\be
 \exp(-\mE r) \rightarrow \exp(- \bar{m}_\rmii{E} r)
 \;, \la{def_bmE}
\ee
where $\bar{m}_\rmii{E}$ contains a logarithmic ultraviolet divergence:
\be
 \bar{m}_\rmii{E,DR} = \mE + 
 \frac{g^2\Nc T }{8\pi}
 \biggl( \frac{1}{\epsilon_\rmii{UV}}  + \ln\frac{\bmu^2}{\mE^2} + \rmO(1) \biggr)
 \;.  \la{bmE_dr}
\ee
A naive transliteration to lattice yields 
\be
 \bar{m}_\rmii{E,lat} = \mE + 
 \frac{g^2\Nc T }{8\pi}
 \biggl( \ln\frac{1}{a^2 \mE^2} + \rmO(1) \biggr)
 \;.  \la{bmE_lat}
\ee
So, Debye screening grows logarithmically as the continuum limit is 
approached, much like in the numerical study of ref.~\cite{lp2}, and  
the exponential function $\exp(- \bar{m}_\rmii{E,lat} r)$ decreases. 

To summarize, we believe that the leading term, 
$\sim \mathcal{G}_\rmii{lat} \exp(-\mE r)$, is a very ultraviolet
sensitive and pathological function of the lattice spacing; at large 
distances, it appears to extrapolate towards zero in the continuum 
limit, while at short distances it appears to explode. 
In fact it might look somewhat like a delta-function. 

Concerning whether the same happens also in the other terms, 
in particular in the one with $\exp(-2\mE r)$, a higher order
computation would be required to see any perturbative indications. 
We find it conceivable, though, that some of the 
terms could also remain finite, representing a coupling to 
the gauge-invariant channel of the traced Polyakov loop,
\be
 \psi_\rmii{T}(r) \equiv
 \frac{1}{\Nc^2}
 \langle \tr[P_\vec{r}] \tr[P^\dagger_\vec{0}] \rangle
 \;. \la{psi_T_def}
\ee
This behaves as~\cite{sn1}
\be
 \ln\biggl( \frac{\psi_\rmii{T} (r)}{|\psi_\rmii{P}|^2} \biggr) \approx 
 \frac{g^4 C_F}{(4\pi)^2 \CA}
 \frac{\exp(-2 \mE r)}{4 T^2 r^2} + \rmO(g^5)
 \la{psi_T_final}
\ee
at the order of our computation; the argument of the large-distance
exponential fall-off has recently been determined at next-to-leading 
order in ref.~\cite{lv}.
(A more precise analysis of this correlator at short
distances has been undertaken in ref.~\cite{average}.)

\begin{figure}[t]


\centerline{%
\epsfysize=8.0cm\epsfbox{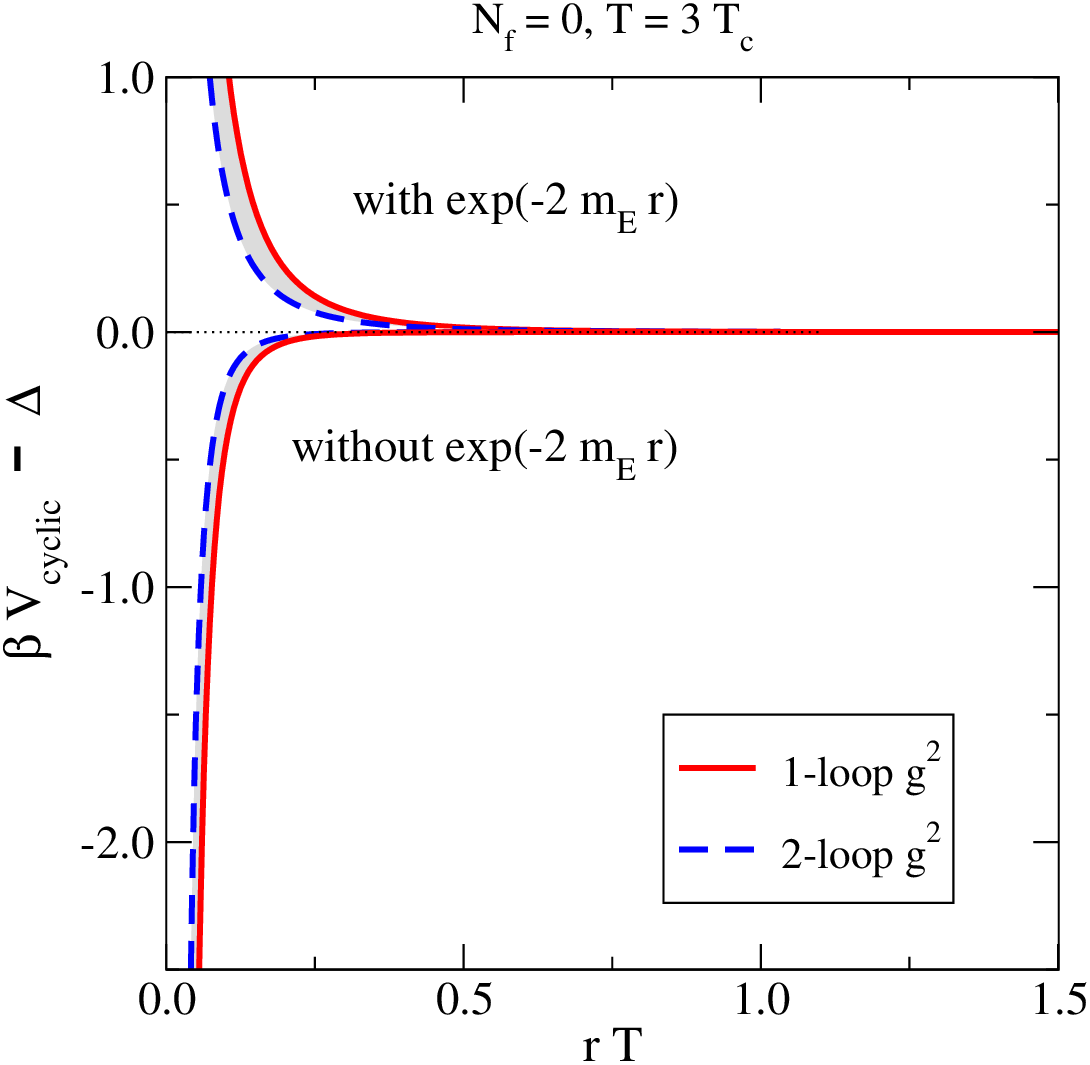}%
}


\caption[a]{\small
Two sketches for the potential extracted from the cyclic Wilson loop, 
$
 \beta V_\rmi{cyclic} \equiv - \ln( {\psi_\rmii{W}} / {|\psi_\rmii{P}|^2})
$, 
after subtracting the strongly cutoff-dependent part 
$\Delta \propto \mathcal{G}_\rmii{DR} \exp(-\mE r)$ from \eq\nr{psi_W_final}. 
The plot is for $\Nf = 0$ and $T = 3.75 \Lambdamsbar$
($T \approx 3 \Tc$). 
The band corresponds to variations of the 
gauge coupling and $\mE$ as explained in the
caption of \fig\ref{fig:psiP}. Comparing with \fig\ref{fig:psiC}(left), 
we see significantly less correlation (stronger screening).
}

\la{fig:psiW}
\end{figure}

A numerical sketch of \eq\nr{psi_W_final}, obtained by omitting 
the regularization dependent leading term altogether, and including
two versions illustrating the importance of the remaining EQCD 
contribution, is given in \fig\ref{fig:psiW}. The main ``prediction''
we can make is that in the continuum limit the result should be much
closer to zero (i.e.\ much more strongly screened) than for the
observable in \fig\ref{fig:Coulomb}. Unfortunately we cannot compare
this prediction with the lattice data from ref.~\cite{bpv}, because
the spatial Wilson lines were smeared in that study. On 
the other hand, a direct lattice measurement of the cyclic Wilson
loop does appear to reproduce both the strong lattice spacing 
dependence as well as the small continuum extrapolation
that are suggested by our analysis~\cite{cyclic_meas}.

%
\section{Conclusions}
\la{se:concl}

The purpose of this paper has been to compute, 
within the weak-coupling expansion in continuum, a number of 
Polyakov loop related observables that have been popular in the context 
of phenomenological studies related to the fate of heavy quarkonium 
at high temperatures. Our main findings can be summarized as follows. 

At large distances, many of the observables considered 
go over to a gauge-independent disconnected contribution, 
the square of the expectation value of a single Polyakov loop. 
We have recomputed this quantity at $\rmO(\alpha_s^2)$ and found
a result different from the classic one in the literature 
(we believe that ours is the correct one). Numerically the 
difference is very small and does not improve on the match to lattice 
data; for that, inclusion of higher order terms 
in the weak-coupling series would probably be needed.

The singlet free energy in Coulomb gauge, defined by \eq\nr{Vsinglet} 
and denoted by $\psi_\rmii{C}$, 
is ultraviolet finite without any additional renormalization factors
in dimensional regularization at order $\rmO(\alpha_s^2)$, 
and agrees at small distances with the gauge-independent zero-temperature
potential~\cite{wf}. At large distances, however, its behaviour 
is not determined by Debye screening, but 
by a power-law contribution, which in our view is related 
to gauge fixing.\footnote{%
 It is interesting to note, though, that ours is not 
 the first study suggesting the existence of power-law terms
 in heavy-quark related observables, cf.\ e.g.\ ref.~\cite{akt}.  
 } 
In principle this means that the Coulomb gauge potential
could be more binding than would physically be expected. On the other hand, 
at intermediate distances the power-law term has no dramatic effect. 
In any case, our result does agree with lattice data surprisingly 
well (cf.\ \fig\ref{fig:psiC}), even at low temperatures; 
a higher-order perturbative computation could perhaps tell whether 
this is an accident or a robust characteristic of thermal 
correlations at non-zero spatial separations.  

The singlet free energy in a general covariant gauge contains
ultraviolet divergences and remains gauge dependent even  
at short distances, where it in general
does not match the zero-temperature potential.

The cyclic Wilson loop is, by construction, extracted from a gauge-invariant
quantity, and indeed the gauge-fixing related power-law term
affecting $\psi_\rmii{C}$ cancels out. 
However, the behaviour of this observable does not match
the zero-temperature potential at short distances. Moreover, 
at large distances
it is not sensitive to physically Debye-screened one-gluon exchange, 
but rather develops a logarithmically ultraviolet 
divergent screening mass. Nevertheless, there could 
remain some non-trivial higher-order $r$-dependence 
in the continuum limit, either with a screening mass 
$\sim 2\mE$ or with $\sim 2\pi T$.

Our findings underline the pitfalls that exist in generic
potential model studies of quarkonium physics at finite temperatures. 
On the other hand, it can be argued that parametrically only the range 
$r T \lsim 1$ is relevant for quarkonium melting~\cite{peskin}, 
so problems that are related to the asymptotic large-distance 
behaviour may not be that important from 
the pragmatic point of view. With this philosophy the Coulomb gauge singlet
free energy seems like the least misleading quantity to use, if it 
is assigned the meaning of the real part of the proper potential~\cite{akr}. 
At the same time, it appears that in this range perturbation theory
alone may yield a fairly good description of the system, so that one
could just as well resort to gauge independent perturbative quarkonium
spectral function determinations such as those in ref.~\cite{nlo}.

%
\section*{Acknowledgements}

We are grateful
to the BMBF for financial support under project
{\em Heavy Quarks as a Bridge between
     Heavy Ion Collisions and QCD}.
M.L.\ was supported in part by the {\em Project of Knowledge
Innovation Program} (PKIP) of the Chinese Academy of Sciences, 
Grant No.\ KJCX2.YW.W10.
M.V.\ was supported by the Academy of Finland, 
contract no.\ 128792. 
We thank 
N.~Brambilla, J.~Ghiglieri, O.~Kaczmarek, P.~Petreczky, 
O.~Philipsen, M.~Tassler and 
A.~Vairo for useful discussions and correspondence, 
and O.~Kaczmarek for providing the lattice 
data for \figs\ref{fig:psiP} and \ref{fig:psiC}.

\appendix
\renewcommand{\thesection}{Appendix~\Alph{section}}
\renewcommand{\thesubsection}{\Alph{section}.\arabic{subsection}}
\renewcommand{\theequation}{\Alph{section}.\arabic{equation}}

%
\section{Sum-integrals and integrals}

In this appendix we list, for completeness, the results for the 
four-dimensional sum-integrals and the three-dimensional
integrals that play a role in our computation.
The theory is regularized dimensionally, with $D = 4 - 2 \epsilon$,
and terms of $\rmO(\epsilon)$ are omitted.
The dependence on the scale parameter, $\bmu$, appears in most cases 
through logarithmic factors which are denoted by $L_\bo, L_\fe$ and
have been defined in \eq\nr{log_defs}. A prime in the sum-integral
symbol indicates that the Matsubara zero mode is left out; when the 
summation-integration variable is in curly brackets, the 
Matsubara frequencies are fermionic.  The four-vector denoted
by $K$ is by definition purely spacelike, 
\be
 K \equiv (0,\vec{k})
 \;. 
\ee
With this notation, 
the following sum-integrals can be determined ($Q^2 \equiv q_n^2 + q^2$): 
\ba
 \Tint{Q} \frac{Q_\mu Q_\nu}{Q^4} & = & 
 - \frac{T^2}{24}
 \Bigl( \delta_{\mu 0}\delta_{\nu 0} - 
 \delta_{\mu i}\delta_{\nu j} \delta_{ij} \Bigr)
 \;, \la{si1} \\ 
 \Tint{\{Q\}} \frac{Q_\mu Q_\nu}{Q^4} & = & 
 \frac{T^2}{48}
 \Bigl( \delta_{\mu 0}\delta_{\nu 0} - 
 \delta_{\mu i}\delta_{\nu j} \delta_{ij} \Bigr)
 \;, \la{si2} \\ 
 \Tint{Q}' \frac{Q_\mu Q_\nu}{Q^6} & = & 
 \frac{1}{(4\pi)^2}
 \biggl\{ 
   \delta_{\mu 0}\delta_{\nu 0}
   \biggl( \frac{1}{4\epsilon} + \frac{L_\bo}{4} + \fr12 \biggr)
   + \delta_{\mu i}\delta_{\nu j} \delta_{ij}
   \biggl( \frac{1}{4\epsilon} + \frac{L_\bo}{4} \biggr)
 \biggr\} 
 \;, \la{si3} \\ 
 \Tint{Q}' \frac{1}{Q^4} & = &  
 \frac{1}{(4\pi)^2}\biggl( \frac{1}{\epsilon} + L_\bo \biggr)
 \;, \la{si4} \\
 \Tint{Q}' \frac{1}{q_n^2 \, Q^2} & = &  
 - \frac{2}{(4\pi)^2}\biggl( \frac{1}{\epsilon} + L_\bo + 2 \biggr)
 \;, \la{si5} \\ 
 \Tint{Q}' \frac{1}{q^2 \, Q^2} & = &  
  \frac{2}{(4\pi)^2}\biggl( \frac{1}{\epsilon} + L_\bo + 2 \biggr)
 \;,  \la{si6}
\ea 
\ba 
 \sum_{q_n'} \int_{\vec{k,q}}
 \frac{1}{q_n^2(q_n^2 + k^2)Q^2} & = & 
 -\frac{1}{(4\pi)^2}
 \;, \la{P_si1} \\
 \sum_{q_n'} \int_{\vec{k,q}}
 \frac{1}{k^2 Q^2(Q+K)^2} & = & 
 -\frac{1}{(4\pi)^2}
 \biggl(
   \frac{1}{4\epsilon} + \ln\frac{\bmu}{2T} + \fr12 
 \biggr)
 \;, \la{P_si2} \\
 \sum_{q_n'} \int_{\vec{k,q}}
 \frac{q_n^2}{k^4 Q^2 (Q+K)^2} & = & 
 \frac{1}{(4\pi)^2}
 \biggl(
   \frac{1}{8} 
 \biggr)
 \;, \la{P_si3} \\
 \sum_{ \{q_n\} } \int_{\vec{k,q}}
 \frac{1}{k^2 Q^2 (Q+K)^2} & = & 
 -\frac{1}{(4\pi)^2}
 \biggl(
   \ln 2
 \biggr)
 \;, \la{P_si4} \\
 \sum_{ \{q_n\} } \int_{\vec{k,q}}
 \frac{q_n^2}{k^4 Q^2 (Q+K)^2} & = & 
 \rmO(\epsilon)
 \;, \la{P_si5} 
\ea
\ba 
  \int_{\vec{k}} \frac{e^{i\vec{k}\cdot\vec{r}}}{k^2}
 \sum_{q_n'} \int_{\vec{q}} \frac{1}{Q^2(Q+K)^2}
  & = &  
 \beta \int_{\vec{k}} \frac{e^{i\vec{k}\cdot\vec{r}}}{k^2}
 \frac{1}{(4\pi)^2}
 \biggl( \frac{1}{\epsilon} + L_\bo \biggr)
 \nn & & \hspace*{-5cm} + \; 
 \frac{1}{8\pi^2}
 \biggl[ 
    \frac{\ln(1-e^{-4\pi T r})}{4\pi T r}
   + \sum_{n=1}^{\infty}
     E_1\Bigl(4\pi T r n\Bigr)
 \biggr] 
 \;, \la{rint_1} \\
  \int_{\vec{k}} \frac{e^{i\vec{k}\cdot\vec{r}}}{k^4}
 \sum_{q_n'} \int_{\vec{q}} \frac{(D-2)q_n^2}{Q^2(Q+K)^2}
  & = & 
 \beta \int_{\vec{k}} \frac{e^{i\vec{k}\cdot\vec{r}}}{k^4}
 \biggl( -\frac{T^2}{12} \biggr)
 + 
 \beta \int_{\vec{k}} \frac{e^{i\vec{k}\cdot\vec{r}}}{k^2}
 \frac{1}{(4\pi)^2}
 \biggl( - \frac{1}{6\epsilon} - \frac{L_\bo}{6} - \fr16 \biggr) \hspace*{5mm}
 \nn & & \hspace*{-5cm} + \; 
 \frac{1}{96\pi^2}
 \biggl[ 
    \frac{-\ln(1-e^{-4\pi T r})}{2\pi T r}
   + \frac{1 + e^{4\pi T r}(4\pi T r - 1)}
 {(e^{4\pi T r}-1)^2}
   - \sum_{n=1}^{\infty} (4\pi T r n)^2
     E_1\Bigl(4\pi T r n\Bigr)
 \biggr] 
 \;, \la{rint_2} \\ 
 \int_{\vec{k}} e^{i\vec{k}\cdot\vec{r}}
 \sum_{q_n'} \int_{\vec{q}} \frac{1}{q^2Q^2(q+k)^2}
  & = & 
 \frac{1}{(4\pi T r)^2}
 \biggl[ 
    \frac{1}{12} - 
    \frac{{\rm Li}_2(e^{-2\pi T r})}{2\pi^2}
 \biggr] 
 \;, \la{rint_3} \\ 
 \int_{\vec{k}} e^{i\vec{k}\cdot\vec{r}}
 \sum_{q_n'} \int_{\vec{q}} \frac{1}{q^2Q^2(Q+K)^2}
  & = & 
 \frac{1}{(4\pi T r)^2}
 \biggl[ 
    \frac{{\rm Li}_2(e^{-2\pi T r})}{2\pi^2}
  - 
    \frac{{\rm Li}_2(e^{-4\pi T r})}{2\pi^2}
 \biggr] 
 \;, \la{rint_4} \\ 
  \int_{\vec{k}} \frac{e^{i\vec{k}\cdot\vec{r}}}{k^2}
 \sum_{\{q_n\}} \int_{\vec{q}} \frac{1}{Q^2(Q+K)^2}
  & = &  
 \beta \int_{\vec{k}} \frac{e^{i\vec{k}\cdot\vec{r}}}{k^2}
 \frac{1}{(4\pi)^2}
 \biggl( \frac{1}{\epsilon} + L_\fe \biggr)
 \nn & & \hspace*{-5cm} + \; 
 \frac{1}{8\pi^2}
 \biggl[ 
    \frac{1}{4\pi T r}\ln\frac{1-e^{-2\pi T r}}{1+e^{-2\pi T r}}
   + \sum_{n=1}^{\infty}
     E_1\Bigl(2 \pi T r \times (2 n - 1) \Bigr)
 \biggr] 
 \;, \la{rint_5} \\  
  \int_{\vec{k}} \frac{e^{i\vec{k}\cdot\vec{r}}}{k^4}
 \sum_{ \{q_n\} } \int_{\vec{q}} \frac{ q_n^2}{Q^2(Q+K)^2}
  & = & 
 \beta \int_{\vec{k}} \frac{e^{i\vec{k}\cdot\vec{r}}}{k^4}
 \biggl( \frac{T^2}{48} \biggr)
 + 
 \beta \int_{\vec{k}} \frac{e^{i\vec{k}\cdot\vec{r}}}{k^2}
 \frac{1}{(4\pi)^2}
 \biggl( - \frac{1}{12\epsilon} - \frac{L_\bo}{12} - \fr16 \biggr)
 \nn & & \hspace*{-5cm} - \; 
 \frac{1}{96\pi^2}
 \biggl[ 
    \frac{1}{4\pi T r}\ln\frac{1-e^{-2\pi T r}}{1+e^{-2\pi T r}}
   + \frac{e^{2\pi T r}}{2}
    \biggl(
      \frac{1}{e^{4\pi T r}-1} - 2 \pi T r 
      \frac{e^{4\pi T r}+1}{(e^{4\pi T r}-1)^2}
    \biggr)
 \nn & & \hspace*{-3.5cm} + \; 
   \fr12 \sum_{n=1}^{\infty} [2\pi T r \times (2n -1) ]^2
     E_1\Bigl(2\pi T r  \times (2 n - 1) \Bigr)
 \biggr] 
 \;, \la{rint_6} 
\ea \ba
  \sum_{q_n'} \int_{\vec{q}}
    \frac{e^{i q_z r} -1}{q_z^2 Q^2} 
  & = &
  \frac{\beta}{8\pi^2} \biggl[ \frac{1}{\epsilon} + L_\bo
  + 2 \ln\Bigl(1-e^{-2\pi T r}\Bigr) 
  + 4 \pi T r \sum_{n=1}^{\infty}
     E_1\Bigl(2\pi T r n\Bigr)
    \biggr]
 \;. \hspace*{1cm} \la{rint_7}
\ea 
Here 
\be
 E_1(z) \equiv 
 \int_z^{\infty} \! {\rm d}t \, \frac{e^{-t}}{t}
 \; ; \qquad
 E_1(z) \stackrel{z \gg 1}{\approx} \frac{e^{-z}}{z}
 \; ; \qquad
 E_1(z) \stackrel{z \ll 1}{\approx} \ln\frac{1}{z} - \gammaE 
 \la{E1}
\ee
is an exponential integral
($E_1(z) = - {\rm Ei}(-z)$ for $z>0$), and 
\be
 {\rm Li}_2(z) \equiv 
 - \int_0^z \! {\rm d}t \, \frac{\ln(1-t)}{t} 
 = 
 \sum_{n=1}^{\infty} \frac{z^n}{n^2}
 \la{Li2}
\ee
is a dilogarithm. 
Where not shown explicitly, the fermionic 
sum-integrals are obtained from the corresponding
bosonic ones by replacing $L_\bo \to L_\fe$.

In \eqs\nr{rint_1}--\nr{rint_7}, various sums over the 
exponential integral appear. All of these can be transformed 
into an integral representation: 
\ba
 \sum_{n=1}^{\infty} E_1(\alpha n) \!\! & = & \!\! 
 -\frac{\ln(1-e^{-\alpha})}{\alpha}
 + \frac{1}{\alpha}
   \int_{1}^{\infty} \frac{{\rm d} x}{x^2} \ln \Bigl( 1- e^{-\alpha x} \Bigr)
 \;, \la{int_rep} \\ 
 \sum_{n=1}^{\infty} (\alpha n)^2 E_1(\alpha n) \!\! & = & \!\!  
 -\frac{2 \ln(1-e^{-\alpha})}{\alpha}
 + \frac{1+ e^{\alpha}(\alpha - 1)}{(e^\alpha - 1)^2}
 + \frac{6}{\alpha}
   \int_{1}^{\infty} \frac{{\rm d} x}{x^4} \ln \Bigl( 1- e^{-\alpha x} \Bigr)
 \;, \qquad \\ 
 \sum_{n\in \NN_\rmii{odd}}  E_1\Bigl(\alpha n \Bigr) \!\! & = & \!\!  
 \frac{1}{2 \alpha}\biggl[ \ln\frac{1+e^{-\alpha}}{1-e^{-\alpha}}
 - \int_{1}^{\infty} \frac{{\rm d} x}{x^2} 
 \ln \frac{1+e^{-\alpha x}}{1-e^{-\alpha x}}
 \biggr]
 \;, \la{int_rep_3} \\ 
 \sum_{n \in \NN_\rmii{odd}} (\alpha n)^2 E_1(\alpha n) \!\! & = & \!\!  
 \frac{1}{\alpha} \ln\frac{1+e^{-\alpha}}{1-e^{-\alpha}}
 + e^\alpha \biggl[ - \frac{1}{e^{2\alpha} - 1}
 + \frac{\alpha (e^{2\alpha} + 1)}{(e^{2\alpha} - 1)^2} \biggr]
 \nn & & 
 \; - \; \frac{3}{\alpha} \int_{1}^{\infty} \frac{{\rm d} x}{x^4} 
 \ln \frac{1+e^{-\alpha x}}{1-e^{-\alpha x}}
 \;. \la{int_rep_last}
\ea

Within EQCD, the following integrals can be worked out:
\ba
 \int_{\vec{k,q}} \frac{1}{(k^2+\mE^2)^2(q^2+\mE^2)}
 & = & 
 \frac{1}{(4\pi)^2}\biggl( -\fr12 \biggr)
 \;, \la{ie1} \\[2mm] 
 \int_{\vec{k,q}} \frac{1}{(k^2+\mE^2) q^2 [(q+k)^2+\mE^2]}
 & = & 
 \frac{1}{(4\pi)^2}\biggl(\frac{1}{4\epsilon}+\ln\frac{\bmu}{2\mE} 
 + \fr12 \biggr)
 \;, \\[2mm] 
 \int_{\vec{k,q}} \frac{\mE^2}{(k^2+\mE^2)^2 q^2 [(q+k)^2+\mE^2]}
 & = & 
 \frac{1}{(4\pi)^2}\biggl( \fr14 \biggr)
 \;. \la{ie3} 
\ea

Finally, we consider the EQCD-integral defined in \eq\nr{I_def}.
In fact, it is useful to re-start with manifestly infrared safe 
integration variables like in \eq\nr{psi_W_full}, and then
the integral can be expressed as
\ba
 \mathcal{I} \!\! 
 & =  & \!\! 
 \int_{\vec{k},\vec{q}} \Bigl( e^{i \vec{k}\cdot\vec{r}} - 
  e^{- i \vec{q}\cdot\vec{r}}\Bigr)
 \frac{q_z - k_z}{q_z + k_z}
     \frac{1}{(q+k)^2 (k^2 + \mE^2)(q^2 + \mE^2)}
 \nn & = &  
 \int_{\vec{k},\vec{q}} e^{i \vec{k}\cdot\vec{r}} \Bigl(1  - 
  e^{i \vec{q}\cdot\vec{r}}\Bigr)
 \frac{q_z + 2 k_z}{q_z} 
     \frac{1}{q^2 (k^2 + \mE^2)[(q+k)^2 + \mE^2]}
 \nn & = &  
 2 \int_\vec{k}\frac{e^{i \vec{k}\cdot\vec{r}}}{k^2 + \mE^2}
    \int_\vec{q} \mathbb{P} \biggl\{ \frac{1}{q_z}
	\frac{q_z + 2 k_z}{q^2 [(q+k)^2 + \mE^2]} \biggr\}
 \nn &=&
 2 \int_\vec{k}\frac{e^{i \vec{k}\cdot\vec{r}}}{k^2 + \mE^2}
    \left[ \int_\vec{q} \frac{1}{q^2 [(q+k)^2 + \mE^2]}
    +\int_\vec{q}\mathbb{P} \biggl\{ \frac{1}{q_z} 
	\frac{2 k_z}{q^2 [(q+k)^2 + \mE^2]} \biggr\} \right]
 \;. \la{i2}
\ea
Here we first substituted $\vec{q} \to -\vec{k} - \vec{q}$;
the numerator vanishes for $q_z = 0$ so the integrand is regular. 
Subsequently we changed integration variables as 
$\vec{k} \to \vec{k} -\vec{q}$ and $\vec{q} \to -\vec{q}$
in the second term, introducing a principal value  
to regulate the potential divergence at $q_z=0$. 
The second term on the last line can be written as
\ba
 \lefteqn{ \int_\vec{q}\mathbb{P} \biggl\{ \frac{1}{q_z} 
 \frac{1}{q^2 [(q+k)^2 + \mE^2]} \biggr\} } 
 \nn &=& \frac{1}{2}\int_\vec{q}  \frac{1}{q_z q^2}
	\left[ \frac{1}{(q+k)_\bot^2 +(k_z+q_z)^2+ \mE^2}
	 - \frac{1}{(q+k)_\bot^2 +(k_z-q_z)^2+ \mE^2} \right] 
 \nn &=& -2k_z \int_\vec{q} 
   \frac{1}{q^2[(q+k)_\bot^2 +(q_z+k_z)^2+ \mE^2]
               [(q+k)_\bot^2 +(q_z-k_z)^2+ \mE^2]}
 \nn &=& -\frac{1}{4\pi} \frac{1}{k^2 + \mE^2} \arctan \frac{k_z}{\mE}
 \;,
\ea
where in the last step the integration could be carried out 
with the help of a Feynman parameter. 
Combining with the first term of the last line of \eq\nr{i2}, we get
\ba
 \mathcal{I} &=&
    2\int_\vec{k}\frac{e^{i \vec{k}\cdot\vec{r}}}{k^2 + \mE^2}
    \left[ \frac{1}{4\pi k} \arctan \frac{k}{\mE} 
	- \frac{1}{4\pi} \frac{2k_z}{k^2 + \mE^2} 
 \arctan \frac{k_z}{\mE} \right]
 \nn &=& \frac{e^{-\mE r}}{16\pi^2 \mE r}
  \Bigl\{ 
          (1+\mE r) \Bigl[ \ln(2\mE r) + \gammaE \Bigr]
        + (1-\mE r) e^{2 \mE r} E_1(2\mE r)
  \Bigr\}
 \;, \la{I_final_2}
\ea
where we proceeded as in connection with \eq\nr{psi_E_final}. 


\section*{Erratum}

\vspace*{-0.95cm}

\hfill (January 2013)

\vspace*{0.5cm}

In the analysis of \se\ref{se:cyclic} several  
Fourier transforms appear 
(\eqs\nr{sd_W_1}, \nr{i1}, \nr{rint_7})
which, taken literally, are ambiguous, 
because of a pole on the integration contour. 
Whereas the first two were defined in the sense of a principal value, 
the last one was defined in the sense of a right value 
(leaving poles to the right of the integration contour). This leads
to an inconsistency when the results are summed together. 
Re-evaluating the last case as a principal value we get 
\begin{eqnarray*}
  \sum_{q_n'} \int_{\vec{q}}
    \mathbbm{P} \biggl( \frac{e^{i q_z r} -1}{q_z^2 Q^2}  \biggr)
  & = &
  \frac{\beta}{8\pi^2} \biggl[ \frac{1}{\epsilon} + L_\bo
  + 2 \ln\Bigl(1-e^{-2\pi T r}\Bigr) 
  + 4 \pi T r \sum_{n=1}^{\infty}
     E_1\Bigl(2\pi T r n\Bigr)
    \biggr]
 \nn 
 & + & 
 \frac{r}{8\pi}
 \biggl[ \frac{1}{\epsilon} + L_\bo + 2 \Bigl( \ln4\pi - \gammaE \Bigr)
 \biggr]
 \;. \hspace*{4.5cm} \hfill (\mbox{A.19}')
\end{eqnarray*} 
The term on the 2nd row, which was absent from the original analysis,
acts as a ``counterterm'' to the long-distance contribution in \eq\nr{i1};
more specifically, 
it cancels the divergences in \eqs\nr{Gdr_large}, \nr{bmE_dr}, 
\nr{bmE_lat}. Consequently, provided that the overall divergences of 
\eqs\nr{Gdr_small}, \nr{Glat_small} can also be successfully taken 
care of, as has recently been proposed in ref.~[47],  
the quantity $\bar{m}_\rmii{E}$ defined in \eq\nr{def_bmE}
could be finite. We thank the authors of ref.~[47] for drawing our 
attention to the inconsistency. 

\bi
 \item[{[47]}]
  M.~Berwein, N.~Brambilla, J.~Ghiglieri and A.~Vairo,
  {\em Renormalization of the cyclic Wilson loop,}
  1212.4413.
\ei

\end{document}